\documentclass[a4paper,oneside,11pt]{article}

\usepackage[english]{babel}
\usepackage[utf8]{inputenc}
\usepackage[T1]{fontenc}

\usepackage[a4paper]{geometry} 
\usepackage{lmodern}
\rmfamily
\DeclareFontShape{T1}{lmr}{b}{sc}{<->ssub*cmr/bx/sc}{}
\DeclareFontShape{T1}{lmr}{bx}{sc}{<->ssub*cmr/bx/sc}{}

\usepackage{graphicx}

\usepackage{subfigure}
\usepackage{subfigmat}

\usepackage[pdftex=true, hyperindex=true, colorlinks=true]{hyperref}

\usepackage{color}
\definecolor{orange}{RGB}{255,165,0}

\usepackage{amsmath}

\usepackage{amsfonts} 
\usepackage{mathtools}
\usepackage{textcomp}
\usepackage{xfrac} 

\usepackage{scrtime} 

\usepackage[fixlanguage]{babelbib}
\selectbiblanguage{english}

\usepackage{caption} 
\usepackage{amssymb}
\usepackage{stmaryrd}

\usepackage{mathptmx}       
\def\du{[\![}
\def\df{]\!]}

\newcommand{\trait}{\rule[0.1cm]{7mm}{0.2mm}\hspace{0.1cm}}

\newcommand{\plm}{\rule[0.1cm]{2mm}{0.2mm}\hspace{2mm}}
\newcommand{\plmf}{\rule[0.1cm]{2mm}{0.2mm}}
\newcommand{\pcm}{\hspace{-0.5mm}\rule[0.1cm]{0.3mm}{0.2mm}\hspace{1.5mm}}
\newcommand{\tirets}{\hspace{0.1cm}\plm\plm\plmf\hspace{0.1cm}}

\newcommand{\dmixte}{\hspace{0.1cm}\plm\pcm\pcm\plm\pcm\pcm\hspace{0.1cm}}

\title{\vspace{-2cm} \textbf{Turbulent jet simulation using high-order DG methods for aeroacoustics analysis}}

\author{
  M. Lorteau\thanks{corresponding author, post-doctoral fellow, DAAA Department, mathieu.lorteau@onera.fr},
  \ M. de la Llave Plata
  \ and V. Couaillier \\
  {\normalsize\itshape
    ONERA The French Aerospace Lab, 92322 Ch\^{a}tillon Cedex, France}\\
}
\date{}

\begin{document}

\maketitle

\renewcommand{\abstractname}{}
\begin{abstract}

\textbf{In this work, a high-order discontinuous Galerkin (DG) method is used to perform a large-eddy simulation (LES) of a subsonic isothermal jet at high Reynolds number $Re_D = 10^6$ on a fully unstructured mesh. Its radiated acoustic field is computed using the Ffowcs Williams and Hawkings formulation. In order to assess the accuracy of the DG method, the simulation results are compared to experimental measurements and a reference simulation based on a finite volume method. The comparisons are made on the flow quantities (mean, rms and spectra) and pressure far field (rms and spectra).}

\end{abstract}

\section{Introduction}

\hspace{0.5cm} Numerical simulations are commonly used as a natural complement to experimental data to analyse in detail the physical process taking place in a turbulent flow. 
Experimental data are limited by the number of probes that can be used and the intrusive character of the measurement technique. The numerical data are, on the other hand, limited by their short relative duration when dealing with scale-resolved simulations. The present work aims at performing a LES of the aerodynamic and acoustic fields of a subsonic turbulent isothermal jet at high Reynolds number ($Re_D = 10^6$).

One of the main difficulties in performing a LES of a turbulent jet consists in having a turbulent jet from the nozzle exit, as in real jets at $Re_D \geq 10^5$ \cite{Husain1979,Zaman1985_1}. To achieve this, a very fine resolution of the boundary layer is required which can be prohibitive in terms of computational cost. This explains why "numerical jets" tend to be laminar right downstream of the nozzle exit. This laminarity of the shear layer has a strong influence on the jet flow development and its radiated acoustic field namely, a shorter potential core length and a stronger turbulence level leading to an overestimation of the acoustic field due to pairings, see \cite{Huet2013,Bogey2011,Bogey2012,Zaman1985_2}.

To date, a number of strategies have been developed to obtain a turbulent flow at nozzle exit. They consist in adding disturbances in the boundary layer inside the nozzle which are either divergence-free "vortex-ring" \cite{Bogey2003,Bogey2005,Lew2005,Uzun2005}, perturbations based on annular mixing layer linear stability theory \cite{Zhao2001,Bodony2005} or a geometrical tripping \cite{Fosso2012,Sanjose2014,Lorteau2015}, in order to force the transition towards a turbulent state. Nonetheless a fine mesh resolution is needed to convect the turbulent fluctuations.

Another approach involves the use of high-order numerical methods such as the discontinuous Galerkin (DG) method \cite{Chapelier2014,Renac2015}. This method has indeed a high order of accuracy, excellent parallel capabilities and can be applied to complex geometries as the formulation is naturally adapted to curvilinear and unstructured meshes. In the DG approach, high order is obtained by increasing the degree of the polynomial basis used to approximate the solution within an element. This is in contrast to finite-volume (FV) methods in which high order is achieved by increasing the size of the numerical stencil. The compactness is thus conserved in DG regardless of the polynomial degree which makes this type of approach ideal for parallel computation. Moreover, the DG method allows for the straightforward implementation of dynamic \emph{hp}-adaptive techniques, see Houston \& Endre~\cite{Houston2005},  Dolej{\v{s}}{\'\i}~\cite{Dolejvsi2013} or Kuru \emph{et al.}~\cite{Kuru2016} for instance. 

In the present work, we have used a modal DG method to perform a LES of a compressible subsonic turbulent jet at Reynolds number $Re_D = 10^6$ and $M = 0.9$. Some simulations have been performed on jet flow with the DG method, see Ham \emph{et al.}~\cite{Ham2009} or Marek \emph{et al.}~\cite{Marek2015} but they have been done at either $2^{nd}$-order or on incompressible jet respectively. To the best of the authors' knowledge, this is one the first high-order simulation of such a flow configuration using a DG approach. The numerical results will be compared to CFD data from a FV code and with the experimental data, available on the same jet configuration.


\section{Simulation parameters}
\label{sec_SimPar}
\subsection{Jet definition}
\label{subsec_JetDef}

\hspace{0.5cm} The nozzle considered in the present paper is a single stream nozzle with a diameter $D_j = 50$~mm. The internal lines of the nozzle are the ones that have been used for the JEAN European project \cite{Jordan2002_1,Jordan2002_2}. An external hood has been added with a 1~mm thick nozzle lip. Figure~\ref{fig_geo_mesh_grid}(a) represents the geometry used for the simulation. The inflow conditions are given at $x/D_j = -5$. Table~\ref{tab_jet_config} summarizes the jet parameters used in the present study.

\begin{table}[h]
 \begin{center}
    \begin{tabular}{*{6}{c}}
  \hline \hline
        $M_j$	& $T_j/T_\infty $	& $T_{tot}/T_\infty $	& $T_\infty (K)$	& $Re_D$		& $p_{tot}/p_\infty$\\ \hline
        0.9		& 1						& 1.158						& 288					& $10^6$	& 1.686771 \\  \hline \hline
  \end{tabular}
\vspace{0.1cm}
    \caption{Jet characteristics with subscripts $j$, $tot$ and $\infty$ respectively representing the flow conditions at the jet axis, jet total properties and ambient conditions}
     \label{tab_jet_config}
 \end{center}
\vspace{-0.5cm}
\end{table}

The jet considered here corresponds to an isothermal subsonic jet at a Mach number $M_j = U_j/c_j = 0.9$  (with $U_j$ and $c_j$ the jet velocity and the sound speed) and a diameter-based Reynolds number $Re_D = U_j.D_j/\nu = 10^6$ (with $\nu$ being the kinematic viscosity). The jet condition corresponds to a static configuration, in which the surrounding flow is considered to be at rest. For numerical stability purposes, an external flow of $U_\infty = 5$~m.s$^{-1}$ is imposed in the simulation, as was done in \cite{Lorteau2015,Lupoglazoff2015,Vuillot2016}. This low velocity of the external flow is not expected to modify the flow characteristics and noise radiation.

\subsection{Numerical method}
\label{subsec_NumMethod}

\hspace{0.5cm} The simulation has been performed based on the LES approach using the Smagorinsky subgrid scale model (with a value $C_S = 0.17$ of the model constant) using the unstructured DG solver {\it Aghora} developed at ONERA \cite{Renac2015,Chapelier2014}. This solver is designed to solve the full set of compressible Navier-Stokes equations, namely,

\vspace{-0.5cm}
\begin{eqnarray}
 \mathbf{U}_t + \nabla \cdot \left( \mathbf{f_c} - \mathbf{f_v} \right) = \mathbf{S} 
 \label{aghora:equations:eqn1}
\end{eqnarray}

\vspace{-0.1cm}
\noindent where $\mathbf{U}=\left(\rho, \rho\mathbf{V}, \rho E \right)^T$ is the vector of conservative variables.  
The vectors $\mathbf{f_c}$, and $\mathbf{f_v}$ are the convective and viscous fluxes, respectively, 
and $\mathbf{S}$ is a source term. 

The DG discretization implemented into {\it Aghora} is based on a modal approach that relies on the use of a hierarchy of orthogonal polynomial functions as basis for the Galerkin projection. A modified Gram-Schmidt orthonormalization procedure ensures the orthogonality of the basis, and thereby the diagonality of the mass matrix \cite{Remacle2003,Bassi2012}. The time integration is performed using an explicit third-order accurate Runge-Kutta method \cite{Gottlieb2001}. The DG method used in this work is briefly outlined below. 

We start by defining a shape-regular partition of the domain $\Omega$, into $N$ non-overlapping and non-empty cells $\kappa$ of characteristic size $h$. We also define the sets ${\cal E}_i$ and ${\cal E}_b$ of interior and boundary faces in $\Omega_h$, such that ${\cal E}_h={\cal E}_i\cup{\cal E}_b$.

Let ${\cal V}_h^p=\{\phi\in L^2(\Omega_h):\;\phi|_{\kappa}\in{\cal P}^p(\kappa),\; \forall\kappa\in\Omega_h\}$ be the functional space formed by tensor products of piecewise polynomials and of either total or partial degree at most $p$, and $(\phi_\kappa^1,\dots,\phi_\kappa^{N_p}) \in {\cal P}^p(\kappa)$ a hierarchical and orthonormal modal basis of ${\cal V}_h^p$, of dimension $N_p$, confined to $\kappa$. The solution in each element is thus expressed as

\vspace{-0.1cm}
\begin{equation}
 {\bf u}_h({\bf x},t)=\sum_{l=1}^{N_p}\phi_\kappa^l({\bf x}){\bf U}_\kappa^{l}(t), \quad \forall{\bf x}\in\kappa,\, \kappa\in\Omega_h,\, \forall t\geq0,
\label{sec:aghora:dg:eqn2}
\end{equation}

\noindent in which the polynomial coefficients $({\bf U}_\kappa^l)_{1\leq l\leq N_p}$ represent the degrees of freedom (DoF) of the discrete problem in element $\kappa$.
The semi-discrete variational form of system of equations (\ref{aghora:equations:eqn1}) thus reads: find ${\bf u}_h$ in ${\cal V}_h^p$ such that $\forall$ $\phi_h$ $\in$ ${\cal V}_h^p$ we have

\vspace{-0.1cm}
\begin{equation}
 \int_{\Omega_h} \phi_h \partial_t {\bf u}_h dV + {\cal L}_c({\bf u}_h,\phi_h) + {\cal L}_v({\bf u}_h,\phi_h)  = \int_{\Omega_h} \phi_h {\bf S}_h 
\label{sec:aghora:dg:eqn1}
\end{equation}

In eqn. (\ref{sec:aghora:dg:eqn1}) ${\cal L}_c$ and ${\cal L}_v$ represent the variational projection of the convective and viscous terms, respectively, onto the functional space ${\cal V}_h^p$. Similarly, the right-hand-side of eqn. (\ref{sec:aghora:dg:eqn1}) is the variational projection of the source term {\bf S} onto ${\cal V}_h^p$.

We now introduce the following notation: for a given interface $e$ in ${\cal E}_i$ we define the average operator $\{u\}=(u^++u^-)/2$ and the jump operator $\du u\df = u^+{\bf n}^+-u^-{\bf n}^-$, where $u^+$ and $u^-$ are the traces of the variable $u$ at the interface between elements $\kappa^+$ and $\kappa^-$. The DG discretization of the convective terms then reads

\vspace{-0.1cm}
\begin{eqnarray}
 {\cal L}_c({\bf u}_h,\phi_h) = &-& \int_{\Omega_h} {\bf f}_c({\bf u}_h)\cdot\nabla_h \phi_h dV + \int_{{\cal E}_i}\du \phi_h\df {\bf h}_c({\bf u}_h^+,{\bf u}_h^-,{\bf n}) dS  + \int_{{\cal E}_b}\phi_h^+ {\bf f}_c\big({\bf u}_b({\bf u}_h^+,{\bf n})\big)\cdot{\bf n} dS\quad \quad
\label{sec:aghora:dg:eqn3}
\end{eqnarray}
The numerical flux ${\bf h}_c$ is chosen so that it is consistent and compact. For the simulations presented in this paper the local Lax-Friedrichs (LLF) flux has been employed. 

The discretization of the viscous terms is performed using the {\it symmetric interior penalty} (SIP) method \cite{Arnold2001}, 

\vspace{-0.1cm}
\begin{eqnarray}
{\cal L}_v({\bf u}_h,\phi_h)= && \int_{\Omega_h} {\bf f}_v({\bf u}_h,\nabla_h{\bf u}_h)\cdot\nabla_h \phi_h dV  
- \int_{{\cal E}_i}\du \phi_h\df \big\{{\bf f}_v\big({\bf u}_h,\nabla_h{\bf u}_h\big)\big\}\cdot{\bf n} dS \nonumber \\
 &&- \int_{{\cal E}_b}\phi_h^+ {\bf f}_v\big({\bf u}_b({\bf u}_h^+,{\bf n}),\nabla{\bf u}_b({\bf u}_h^+,{\bf n})\big)\cdot{\bf n} dS 
- \int_{{\cal E}_i}\du{\bf u}_h\df \big\{{\bf G}^T\big({\bf u}_h\big)\nabla_h \phi_h\big\}\cdot{\bf n} dS  \nonumber \\
&&- \int_{{\cal E}_b}({\bf u}_h^+ - {\bf u}_b({\bf u}_h^+,{\bf n})) \big\{{\bf G}^T\big({\bf u}_b\big)\nabla_h \phi_h^+\big)\big\}\cdot{\bf n} dS 
+ \int_{{\cal E}_i}\eta_{IP}\du{\bf u}_h\df \du\phi_h\df  dS \nonumber \\
&&+ \int_{{\cal E}_b}\eta_{IP}\big({\bf u}_h^+-{\bf u}_b({\bf u}_h^+,{\bf n})\big)\phi_h^+ dS  
\label{sec:aghora:dg:eqn4}
\end{eqnarray}

\noindent where ${\bf G}= \partial {\bf f}_v/\partial \big(\nabla_h{\bf u}_h\big)$ is so-called homogeneity tensor. The penalty parameter $\eta_{IP}$ has to be chosen sufficiently large to ensure the coercivity of the bilinear form and thus the numerical stability of the simulation \cite{Arnold2001}.

\subsection{Grid parameters and boundary conditions}
\label{subsec_GridPar_BC}

\begin{figure}[h!]
	\begin{center}
		\begin{minipage}[c]{0.49\linewidth}
			\centering \includegraphics[height=4cm]{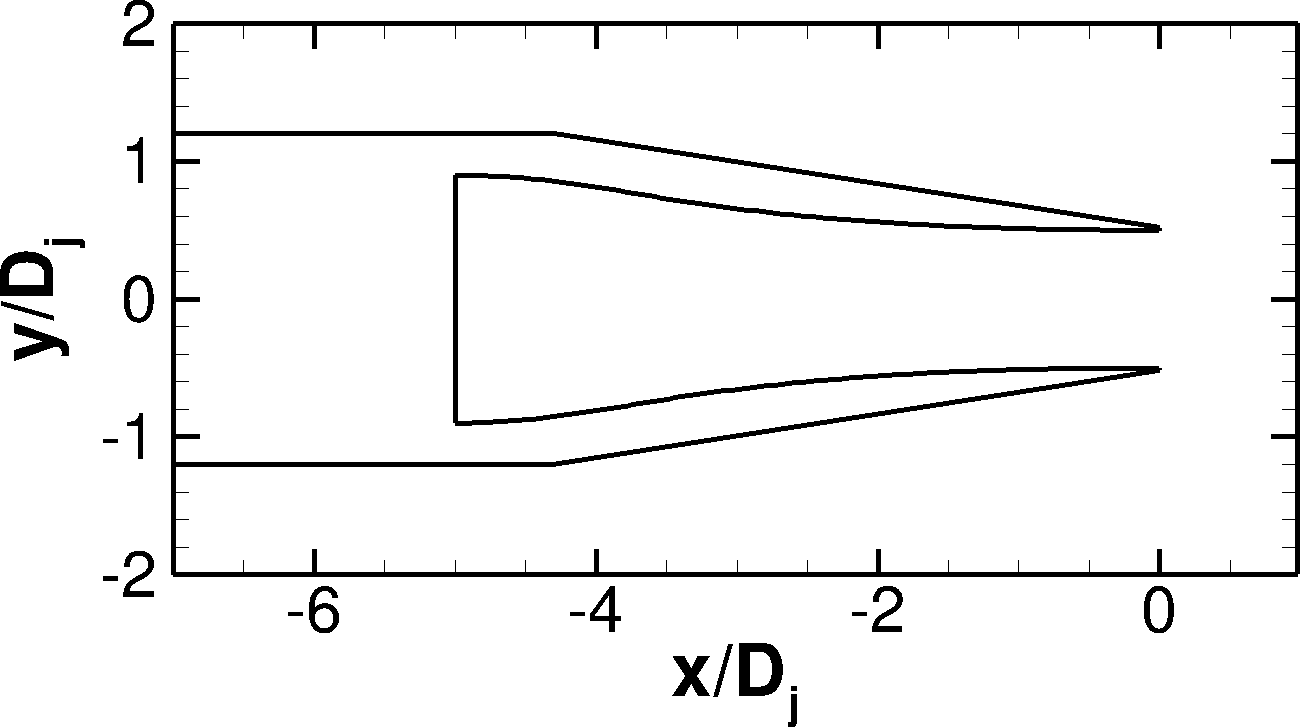}
		\end{minipage}
		\hfill
		\begin{minipage}[c]{0.49\linewidth}
			\centering \includegraphics[height=5cm]{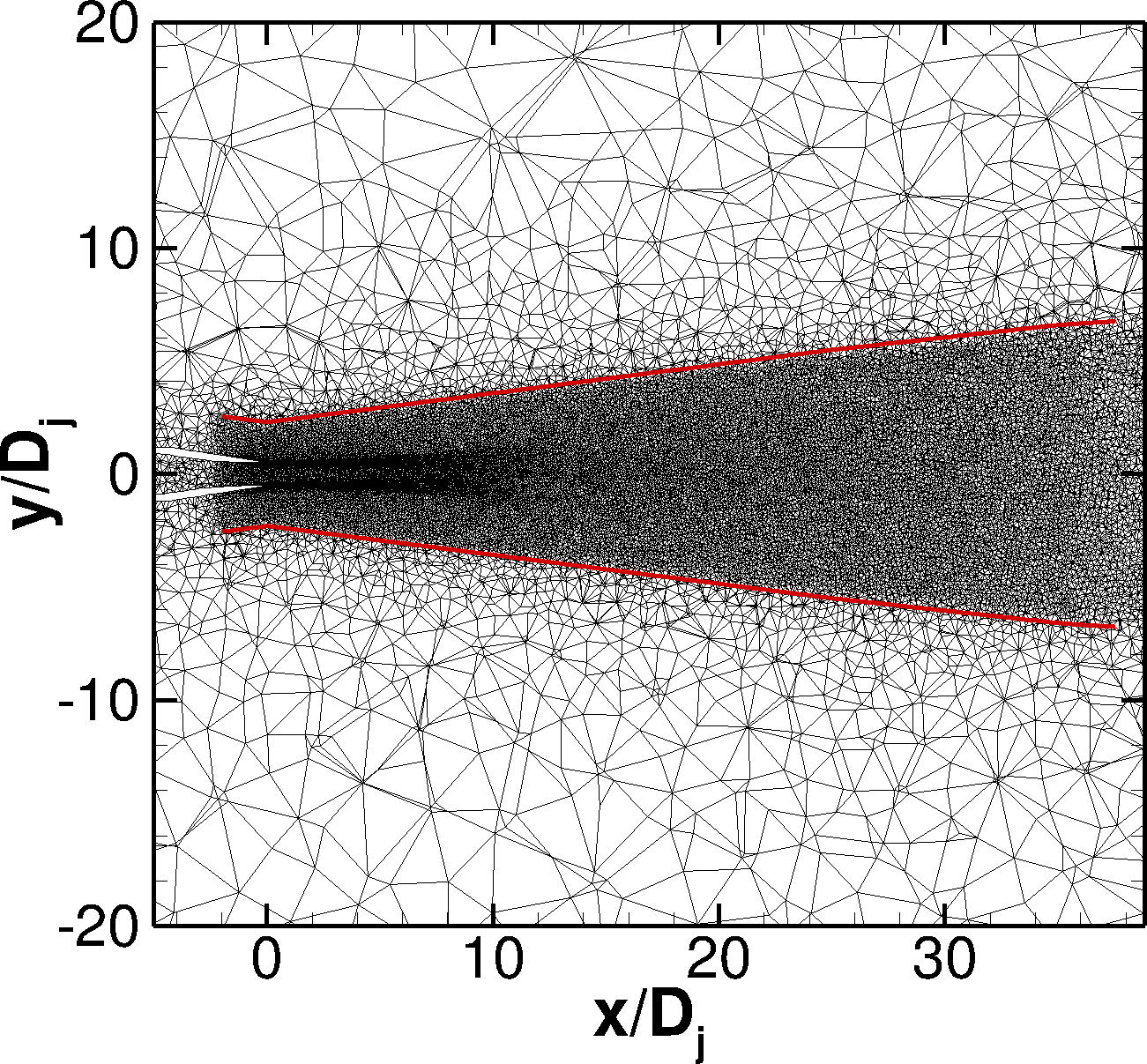}
		\end{minipage}
		\hfill
		\begin{minipage}[c]{0.49\linewidth}
			\vspace{0.1cm}
			\centering (a) $\phi50$ nozzle geometry
		\end{minipage}
		\hfill
		\begin{minipage}[c]{0.49\linewidth}
			\vspace{0.1cm}
			\centering (b) global view of the mesh
		\end{minipage}
		\hfill
		\begin{minipage}[c]{\linewidth}
			\vspace{0.1cm}
	 		\caption{View of the nozzle geometry and of the grid. The red line highlights the refined mesh zone.}
 			\label{fig_geo_mesh_grid}
		\end{minipage}	
	\end{center}
\vspace{-0.5cm}
\end{figure}

\hspace{0.5cm} The computational domain is cylindrical with a radius of $80D_j$ and extends from $x/D_j = -20$ to $x/D_j = 100$. The grid used is second order and fully unstructured. A refined mesh zone which extends downstream of the nozzle exit from ($x/D_j = 0$; $r/D_j = 2.3$) to $x/D_j = 35$; $r/D_j = 6.6$) and upstream of the nozzle exit up to ($x/D_j = -2$; $r/D_j = 2.55$) ensures a good resolution of the jet flow development area. This zone, shown in figure~\ref{fig_geo_mesh_grid}(b), contains the majority of the cells (around $86\%$) and inside, the maximum mesh size is $\Delta x_{max} = 7.6$~mm. This enables to resolve acoustic waves up to a Strouhal number $St = f.D_j/U_j = 1.5$ with 5 cells per wavelength for the polynomial degree $p = 3$ and the third-order time-marching scheme considered (see section~\ref{subsubsec_CAA}).

The meshing methodology employed here is described in \cite{Lorteau2015,Lupoglazoff2015} for the jet noise simulation. This methodology consists in having a fine resolution of the mesh near the nozzle exit. In particular, a fine azimuthal resolution and a moderate mesh stretching were employed to capture the turbulent structures. 

\begin{figure}[h!]
	\begin{center}
		\begin{minipage}[c]{\linewidth}
			\centering \includegraphics[height=4cm]{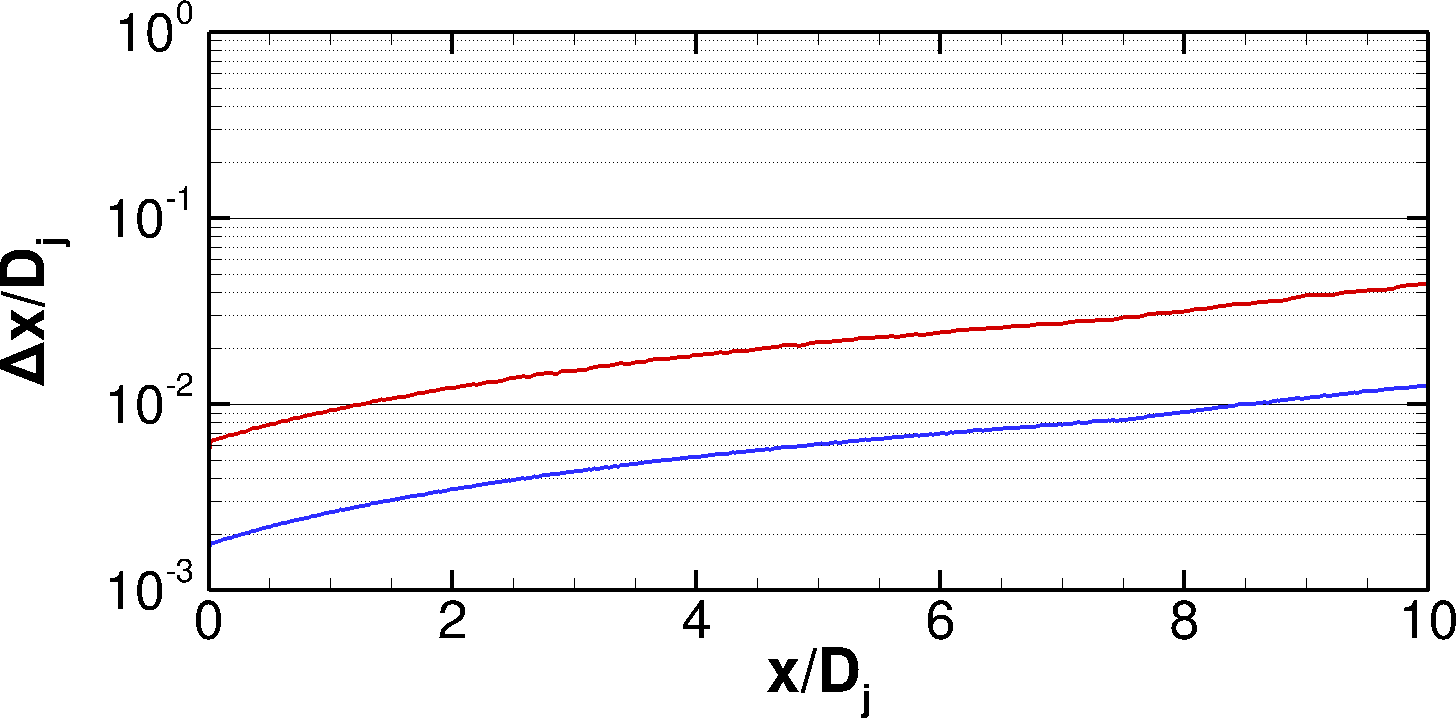}
		\end{minipage}
		\hfill
		\begin{minipage}[c]{\linewidth}
			\vspace{0.1cm}
	 		\caption[]{Comparison of the longitudinal evolution of the mesh size at $r/D_j = 0.5$. \textcolor{red}{\trait}, \emph{Aghora} DGp3 simulation; \textcolor{blue}{\trait}, CEDRE FV2 simulation}
 			\label{fig_taille_maille}
		\end{minipage}	
	\end{center}
\vspace{-0.5cm}
\end{figure}

At the nozzle lip, the cells are isotropic which is well suited for isotropic turbulence, shown to take place in the jet shear layer \cite{Fleury2008}. The cell size $\Delta x/D_j = 0.8\%$ leads to approximately 390 cells in the azimuthal direction.
Taking into account these considerations, we obtain a mesh composed of $3.9\times10^6$ tetrahedra in total. This is less than what has been used for the aforementioned finite volume simulations \cite{Lorteau2015,Lupoglazoff2015,Vuillot2016}. However, the present DG simulation performed with higher order of accuracy appears to be less restrictive in terms of cells number, compared to a finite volume method. This is specially so in terms of the number of degrees of freedom (DoFs) \cite{Chapelier2014}. The longitudinal cell size in the shear layer (i.e.\ at $r/D_j = 0.5$) is shown in figure~\ref{fig_taille_maille} and the mesh characteristics are summarized in table~\ref{tab_mesh_charac}.


In the following, the \emph{Aghora} DGP3 simulation is compared to the reference finite volume CEDRE FV2 simulation performed on the same jet configuration. Details of the CEDRE FV2 simulation are also presented in figure~\ref{fig_taille_maille} and table~\ref{tab_mesh_charac}. The CEDRE FV2 simulation has been performed using the CEDRE code developed at ONERA \cite{Refloch2011}. CEDRE is a multi-physics solver which has been developed for industrial and research applications in the fields of energetics and propulsion and has been used for jet noise simulation. The resolution of the Navier-Stokes equations is based on a finite volume approach for the conservative variables on generalized unstructured meshes. This simulation has been performed using 2$^{nd}$-order space and time schemes on a tetrahedral mesh composed of $165\times10^6$ cells generated using the above mentioned meshing methodology \cite{Lorteau2015,Lupoglazoff2015}. Such 2$^{nd}$-order FV simulations have been validated in previous work and will thus serve as a reference simulation to assess the quality of the DG solution. The mesh used in the \emph{Aghora} simulation has been generated based on the mesh used in the CEDRE simulation (with coarser cells). Thus the two simulations are performed using similar meshes as regards to the refined regions and topology. 
\\


\begin{table}[h]
 	\begin{center}
    	\begin{tabular}{l*{6}{c}}
  		\hline \hline
      Grid name						& $\Delta/D_j$ (\%)	& $n_\theta$	& $St_{cut-off} $	& \#cells ($\times 10^6$)	& \#DoFs ($\times 10^6$)	& $TU_j/D_j$\\ \hline
      \emph{Aghora} DGp3	& 0.8						& 390				& 1.5						& 3.9									& 78												& 180 \\  \hline
      CEDRE FV2					& 0.38						& 838				& 1.0						& 165									& 165											& 250 \\  \hline \hline
  		\end{tabular}
		\vspace{0.1cm}
    	\caption{Characteristics of the different grids, the mesh size $\Delta/D_j$ and the number of cells in the azimuthal direction $n_\theta$ are given near $x/D_j = 0$ and $r/D_j = 0.5$. $T$ represents the simulated duration.}
     	\label{tab_mesh_charac}
 	\end{center}
	\vspace{-0.5cm}
\end{table}

At the nozzle inlet, uniform profiles of stagnation pressure and temperature are imposed as well as an axial velocity direction. The boundary layer thus develops freely. Outside the nozzle, a static pressure $p_\infty = 100310$~Pa is imposed at the outflow boundary. On the lateral and upstream boundaries, a non-reflecting boundary condition is imposed using an external reference state with $p_\infty = 100310$~Pa, $T_\infty = 288$~K and an axial velocity $U_\infty = 5$~m.s$^{-1}$. The grid is stretched from a refined zone, in which the flow is accurately calculated, to the external boundaries of the computational domain. This avoids any spurious reflections by damping the acoustic waves before they reach the borders. The nozzle walls are assumed to be adiabatic.

\subsection{Acoustic computation}
\label{subsec_FWH}

\hspace{0.5cm} The noise radiation is performed using the Ffowcs Williams \& Hawkings \cite{FWH1969} porous surface formulation available	in the code KIM developed at ONERA \cite{Rahier2004}. This formulation allows us to compute pressure time histories at any observer location by integration of the flow field solution on a control surface surrounding the jet and containing all the noise sources. Here, the surface used is represented by the red line in figure~\ref{fig_geo_mesh_grid}(b).


\section{Validation}
\label{sec_Valid}

%


\hspace{0.5cm} In this section, we present aerodynamic and acoustic results from the \emph{Aghora} DGP3 simulation with comparisons to measurements in order to see how the present simulation reproduces the jet configuration. The results are also compared with the aforementioned CEDRE FV2 simulation. 


In the following, we first study the shear layer development and pay special attention to the initial state of the shear layer. Then the jet development is examined and finally, we compare the far pressure field obtained from the DG simulation to the experimental data.  


\subsection{Shear layer}
\label{subsec_ShearLayer}

\begin{figure}[h!]
	\begin{center}
		\begin{minipage}[c]{0.49\linewidth}
			\centering \includegraphics[height=3.5cm]{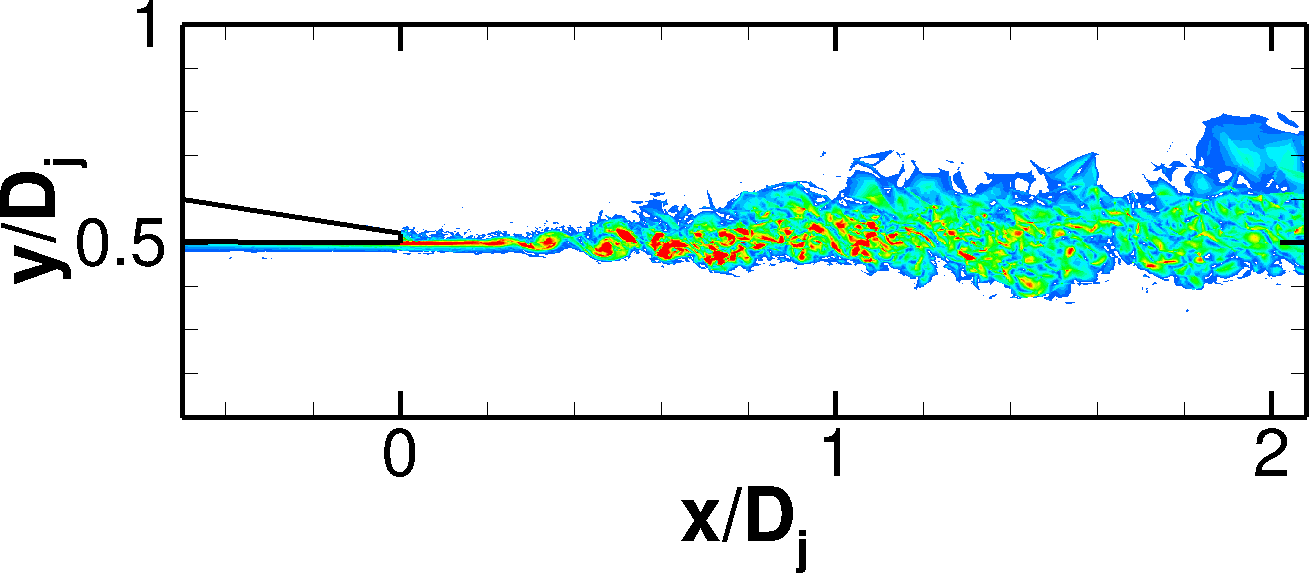}
		\end{minipage}
		\hfill
		\begin{minipage}[c]{0.49\linewidth}
			\centering \includegraphics[height=3.5cm]{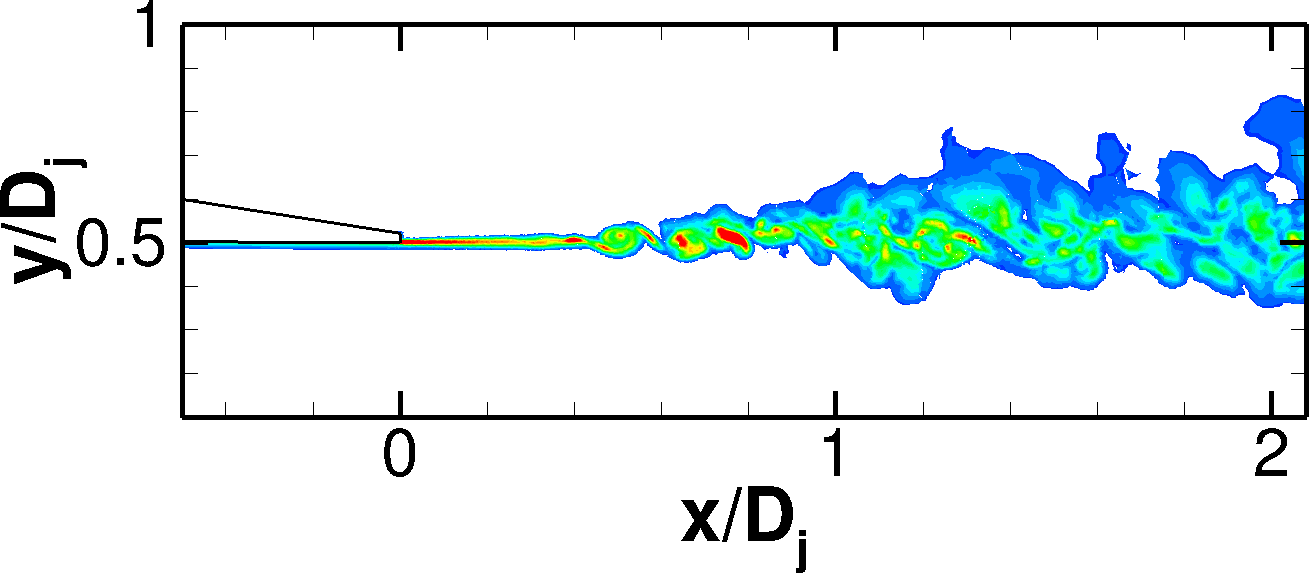}
		\end{minipage}
		\hfill
		\vspace{0.2cm}
		\begin{minipage}[c]{0.49\linewidth}
			\centering (a) \emph{Aghora} simulation
		\end{minipage}
		\hfill
		\vspace{0.2cm}
		\begin{minipage}[c]{0.49\linewidth}
			\centering (b) CEDRE simulation
		\end{minipage}
		\hfill
		\begin{minipage}[c]{\linewidth}
			\vspace{0.1cm}
	 		\caption[]{Snapshots in the (x,r) plane of the vorticity norm in the boundary layer and the shear layer downstream of the lip line for $0 \leq  |\omega| \leq 60U_j/D_j$}
 			\label{fig_vort_sortie_tuyere_COMP_aghora_cedre}
		\end{minipage}	
	\end{center}
\vspace{-0.8cm}
\end{figure}

\hspace{0.5cm} Figure~\ref{fig_vort_sortie_tuyere_COMP_aghora_cedre} reproduces a snapshot of the field of unsteady vorticity magnitude $|\omega|$ at the nozzle exit for the two simulations. It appears that in both simulations a part of the shear layer is laminar and exhibits pairings downstream of the nozzle exit. The transition of the shear layer towards a turbulent state seems to occur at $x/D_j = 0.4$.
\\

\begin{figure}[h!]
	\begin{center}
		\begin{minipage}[c]{\linewidth}
			\centering \includegraphics[height=4cm]{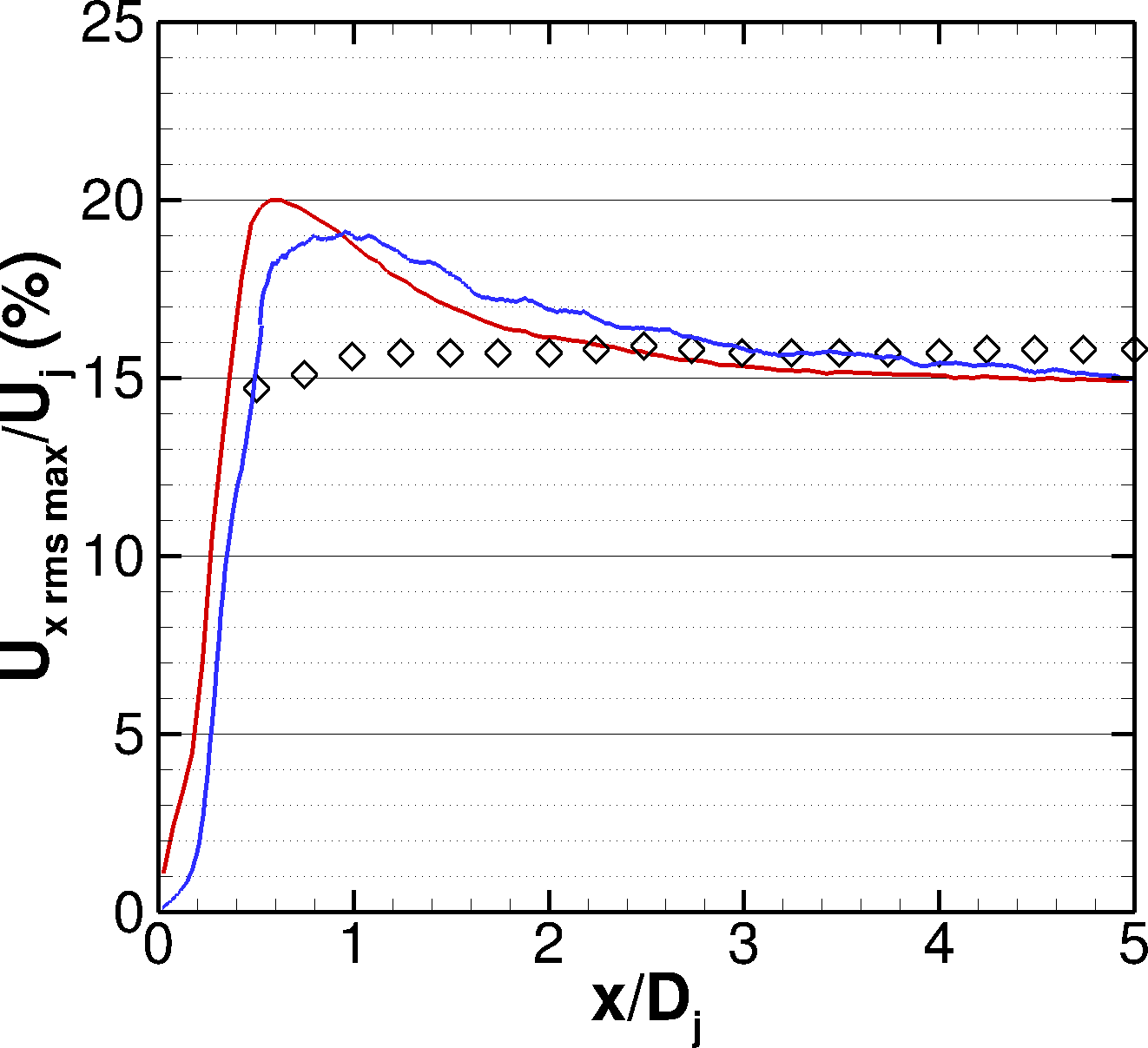}
		\end{minipage}
		\hfill
		\begin{minipage}[c]{\linewidth}
			\vspace{0.1cm}
	 		\caption[]{Comparison of the longitudinal evolution of the peak rms value of axial velocity. \textcolor{red}{\trait}, \emph{Aghora} simulation; \textcolor{blue}{\trait}, CEDRE simulation; $\diamond$, Fleury \emph{et al.} ($M=0.9$, $Re=7.7\times10^5$)}
 			\label{fig_Ux_rms_max_aghora_cedre}
		\end{minipage}	
	\end{center}
\vspace{-0.8cm}
\end{figure}

We see in figure~\ref{fig_Ux_rms_max_aghora_cedre} that the two simulations have a different longitudinal evolution of the peak of the rms axial velocity close to the nozzle exit for $x/D_j \leq 2$. The \emph{Aghora} simulation has an initial peak value of around $1\%$. Right after the nozzle exit ($x/D_j \leq 0.6$), the peak value of the rms axial velocity has a strong growth up to a peak value of around $20\%$ and after decreases to attain a value of around $15\%$.  This behaviour is quite similar to the one observed by Bogey \emph{et al.} \cite{Bogey2012} in a simulation with a low initial turbulence level. However Bogey \emph{et al.} showed that an initial turbulence level of at least $9\%$ is required to have a jet which is initially turbulent. According to this criterion, the jet for the \emph{Aghora} simulation is not totally turbulent. The CEDRE simulation presents a similar evolution as the \emph{Aghora} simulation but the transition to turbulence appears to be less pronounced than in the \emph{Aghora} simulation. Indeed, the peak value is smaller ($\sim 19\%$) and is reached downstream to the position reached for the \emph{Aghora} simulation ($x/D_j \approx 0.8$ against 0.6). 
\\

\begin{figure}[h!]
	\begin{center}
		\begin{minipage}[c]{0.32\linewidth}
			\centering \includegraphics[height=4cm]{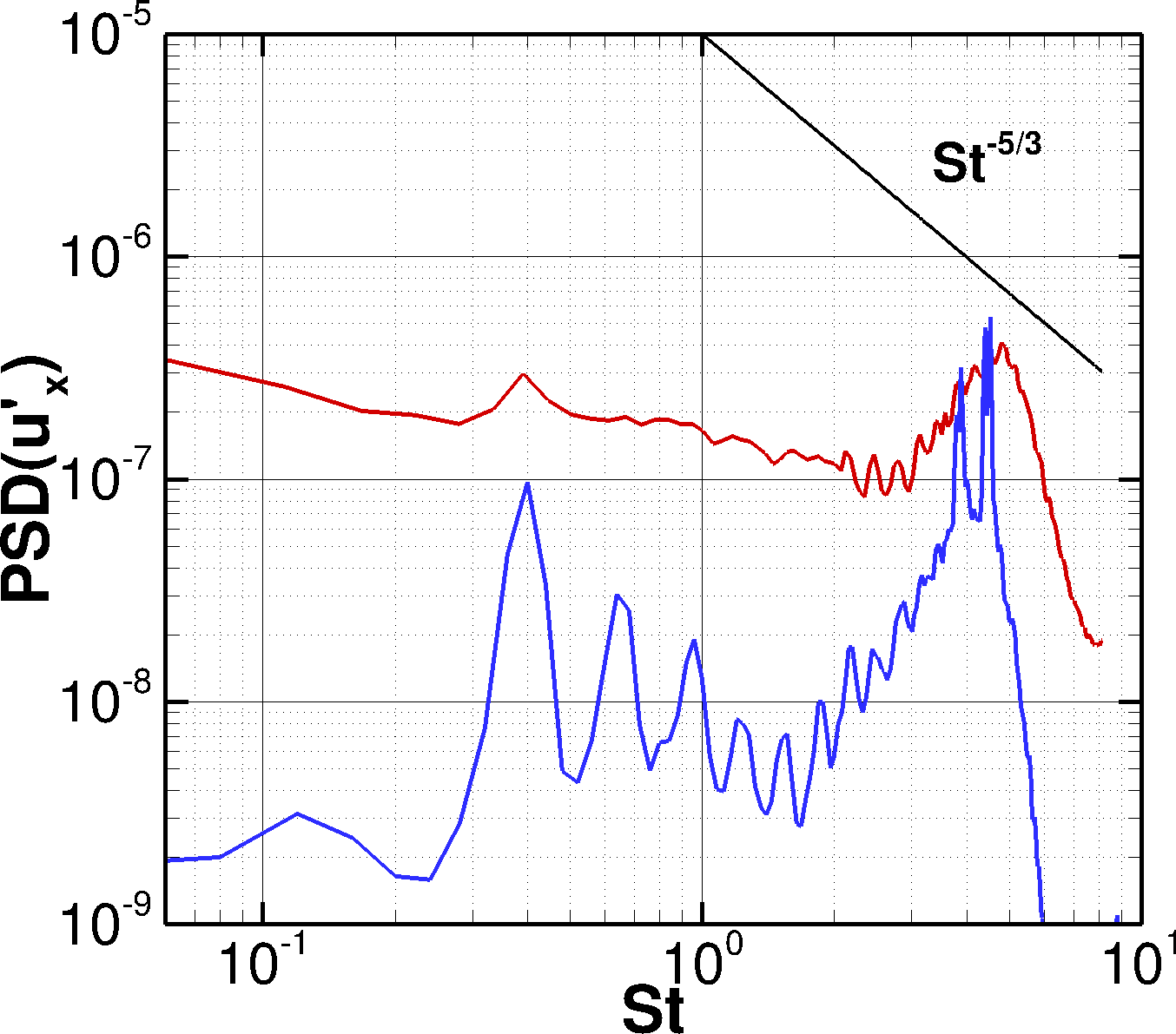}
		\end{minipage}
		\hfill
		\begin{minipage}[c]{0.32\linewidth}
			\centering \includegraphics[height=4cm]{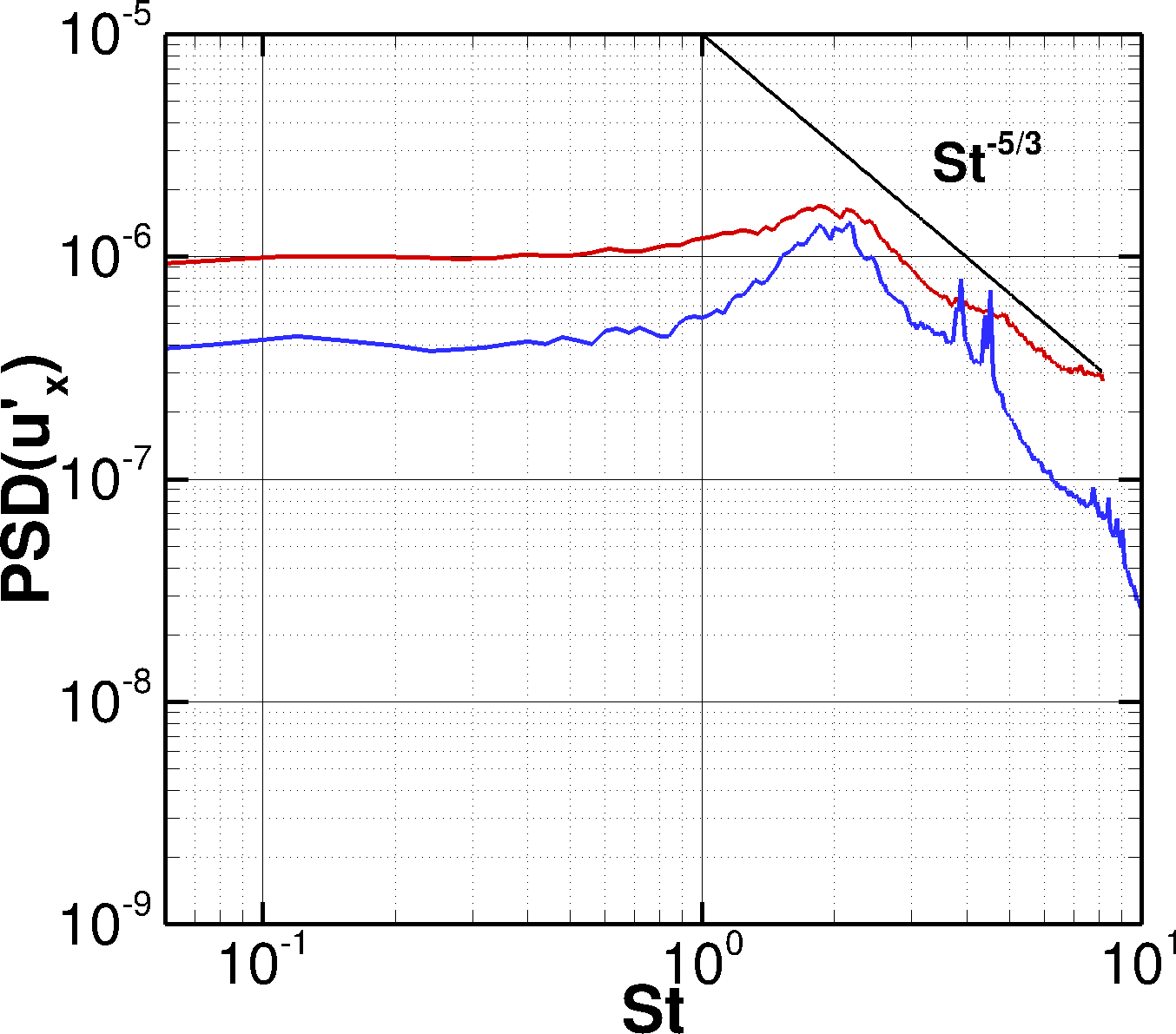}
		\end{minipage}
		\hfill
		\begin{minipage}[c]{0.32\linewidth}
			\centering \includegraphics[height=4cm]{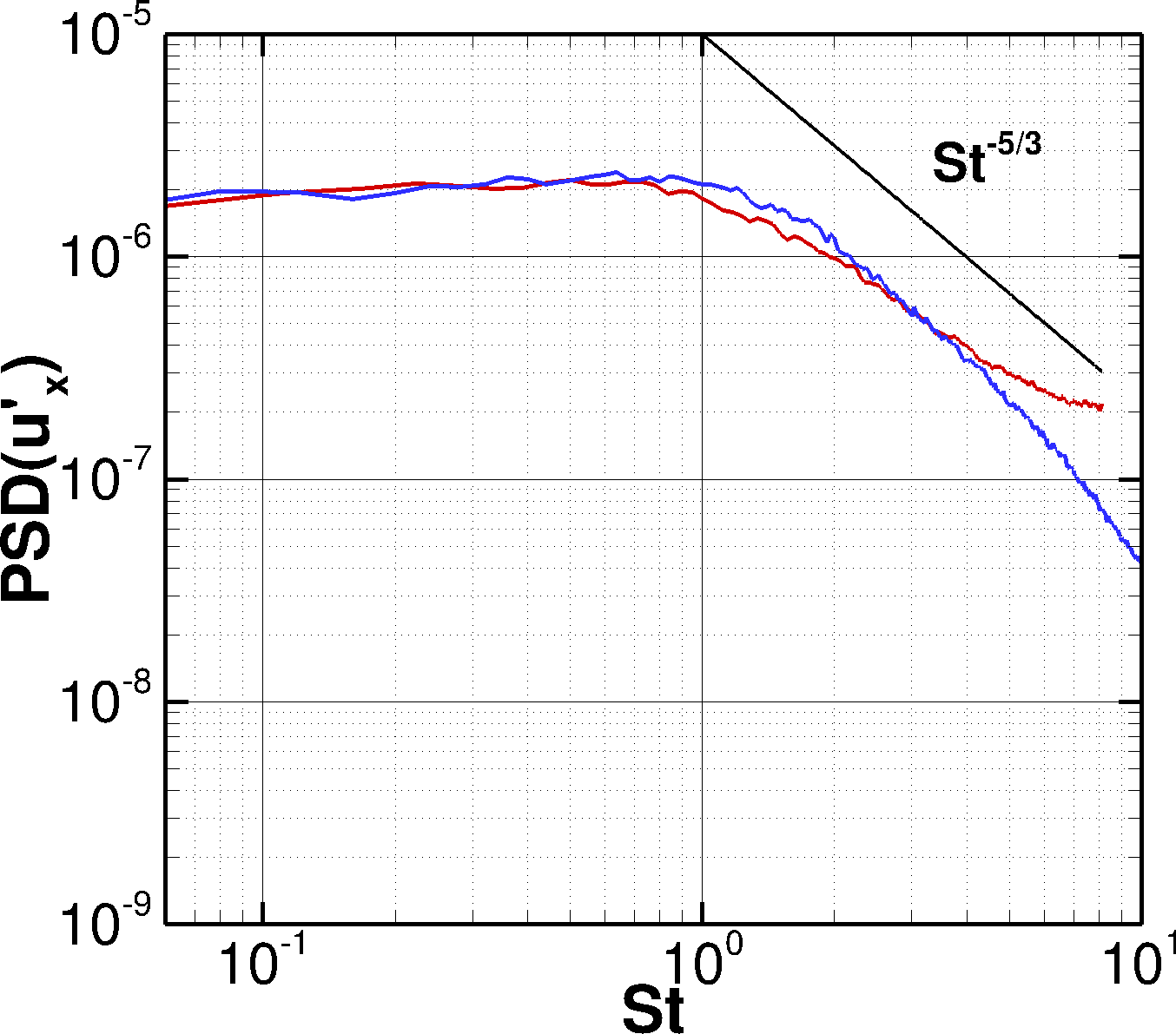}
		\end{minipage}
		\hfill
		\vspace{0.1cm}
		\begin{minipage}[c]{0.32\linewidth}
			\centering (a) $x/D_j = 0.25$
		\end{minipage}
		\hfill
		\vspace{0.1cm}
		\begin{minipage}[c]{0.32\linewidth}
			\centering (b) $x/D_j = 0.5$
		\end{minipage}
		\hfill
		\vspace{0.1cm}
		\begin{minipage}[c]{0.32\linewidth}
			\centering (c) $x/D_j = 1$
		\end{minipage}
		\hfill
		\begin{minipage}[c]{\linewidth}
			\vspace{0.1cm}
	 		\caption[]{Comparison of PSD of axial fluctuating velocity adimensioned by $U_j$ at $r/D_j = 0.5$ for $x/D_j \in \{0.25; 0.5; 1\}$. \textcolor{red}{\trait}, \emph{Aghora} simulation; \textcolor{blue}{\trait}, CEDRE simulation}
 			\label{fig_DSP_Ux_cisail_COMP_aghora_cedre}
		\end{minipage}	
	\end{center}
\vspace{-0.8cm}
\end{figure}

Figure~\ref{fig_DSP_Ux_cisail_COMP_aghora_cedre} presents the power spectral density (PSD) of the axial fluctuating velocity at different axial positions close to the nozzle in the shear layer. For both simulations the PSDs exhibit several peaks at $St \approx 4$ for $x/D_j = 0.25$ and $St \approx 2$ for $x/D_j = 0.5$ which are less and less pronounced as we move downstream. The \emph{Aghora} solution has nonetheless less marked peaks than the CEDRE solution. This might be linked to the higher level of turbulence at these positions in the \emph{Aghora} simulation (see figure~\ref{fig_Ux_rms_max_aghora_cedre}). These peaks are related to the initially laminar state of the shear layer. For the positions downstream of $x/D_j = 1$, the spectra are rather flat and start to decrease for $St \geq 1$ according to a $St^{-5/3}$ law.
\\

\begin{figure}[h!]
	\begin{center}
		\begin{minipage}[c]{0.32\linewidth}
			\centering \includegraphics[height=4cm]{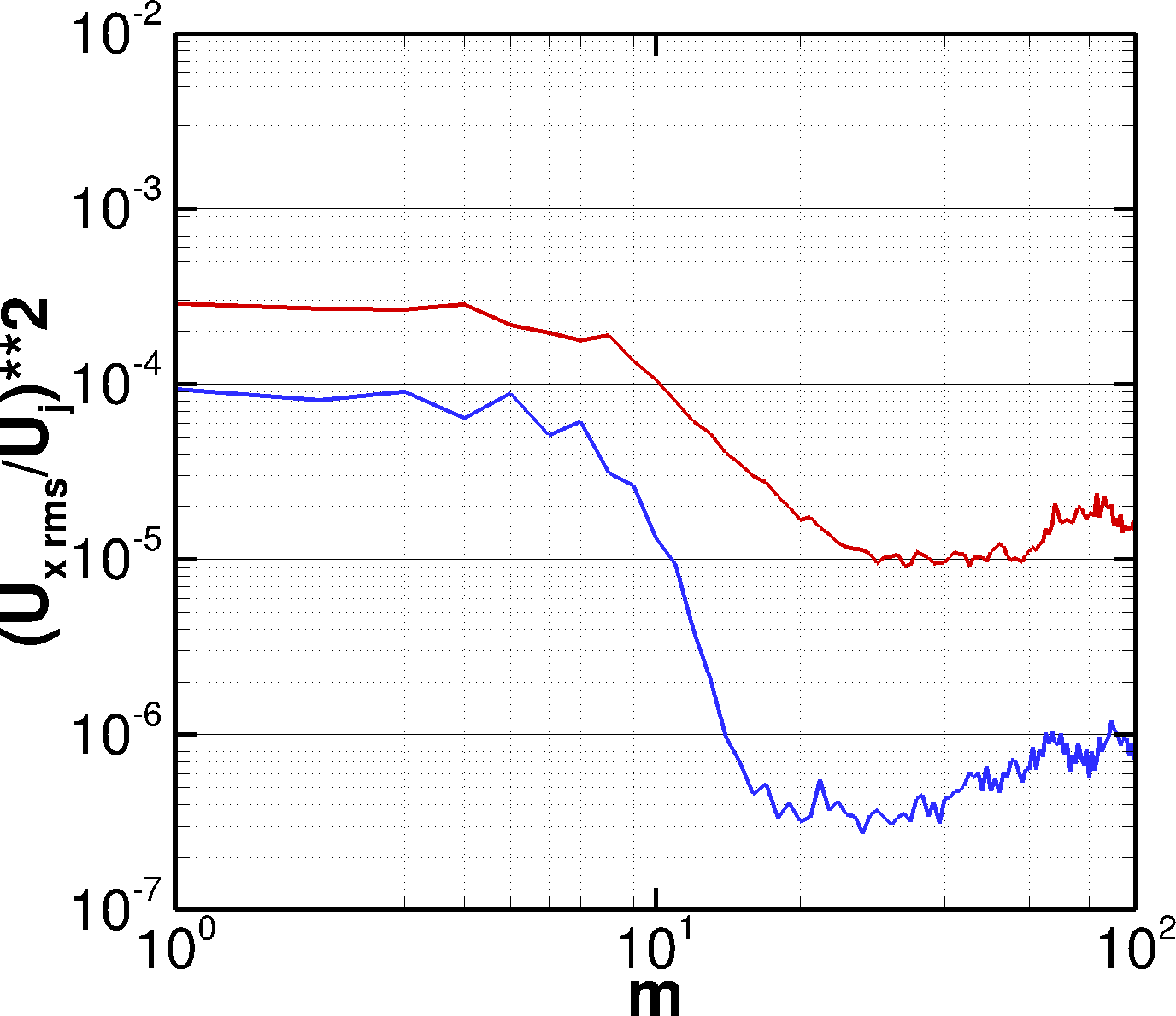}
		\end{minipage}
		\hfill
		\begin{minipage}[c]{0.32\linewidth}
			\centering \includegraphics[height=4cm]{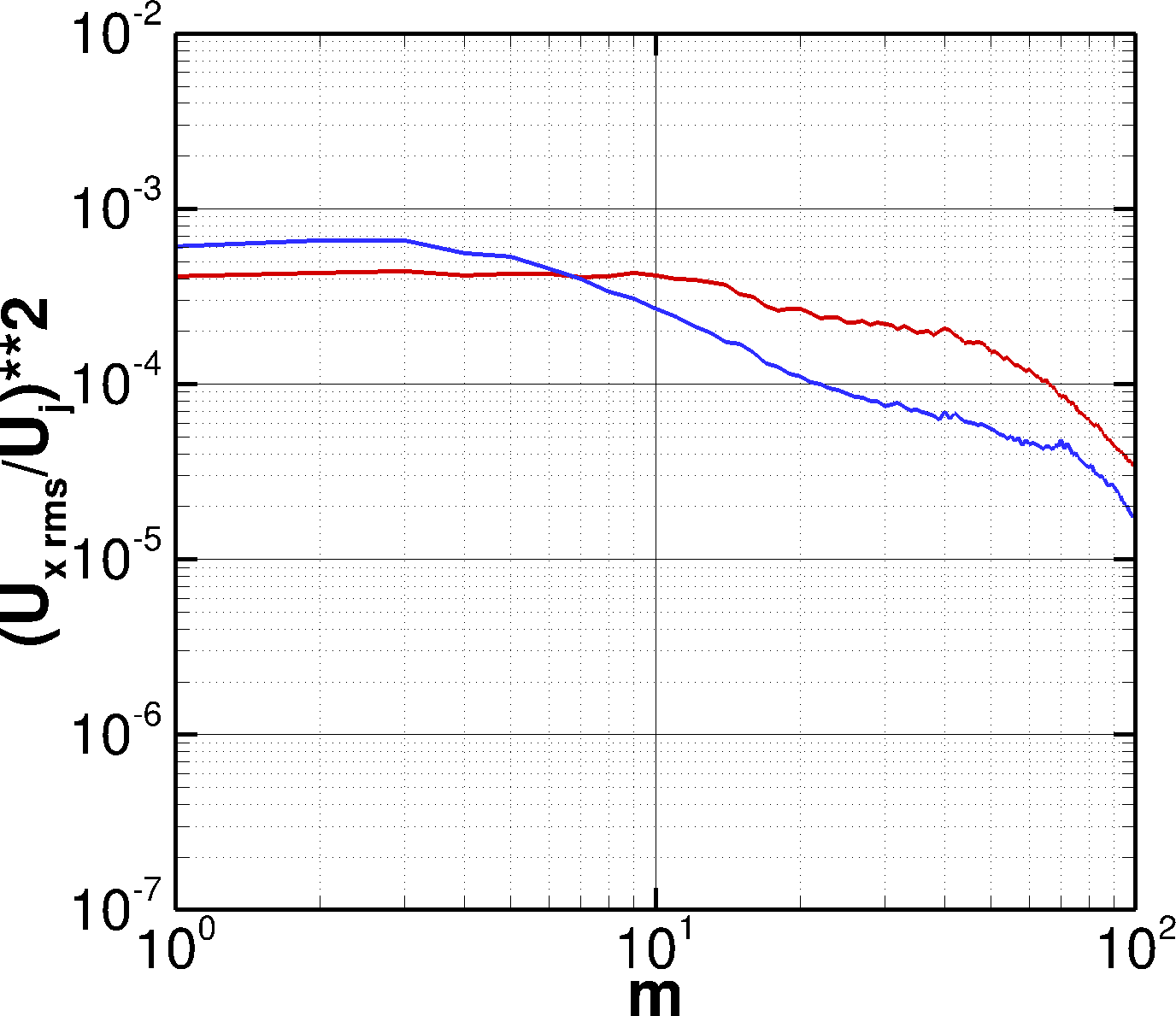}
		\end{minipage}
		\hfill
		\begin{minipage}[c]{0.32\linewidth}
			\centering \includegraphics[height=4cm]{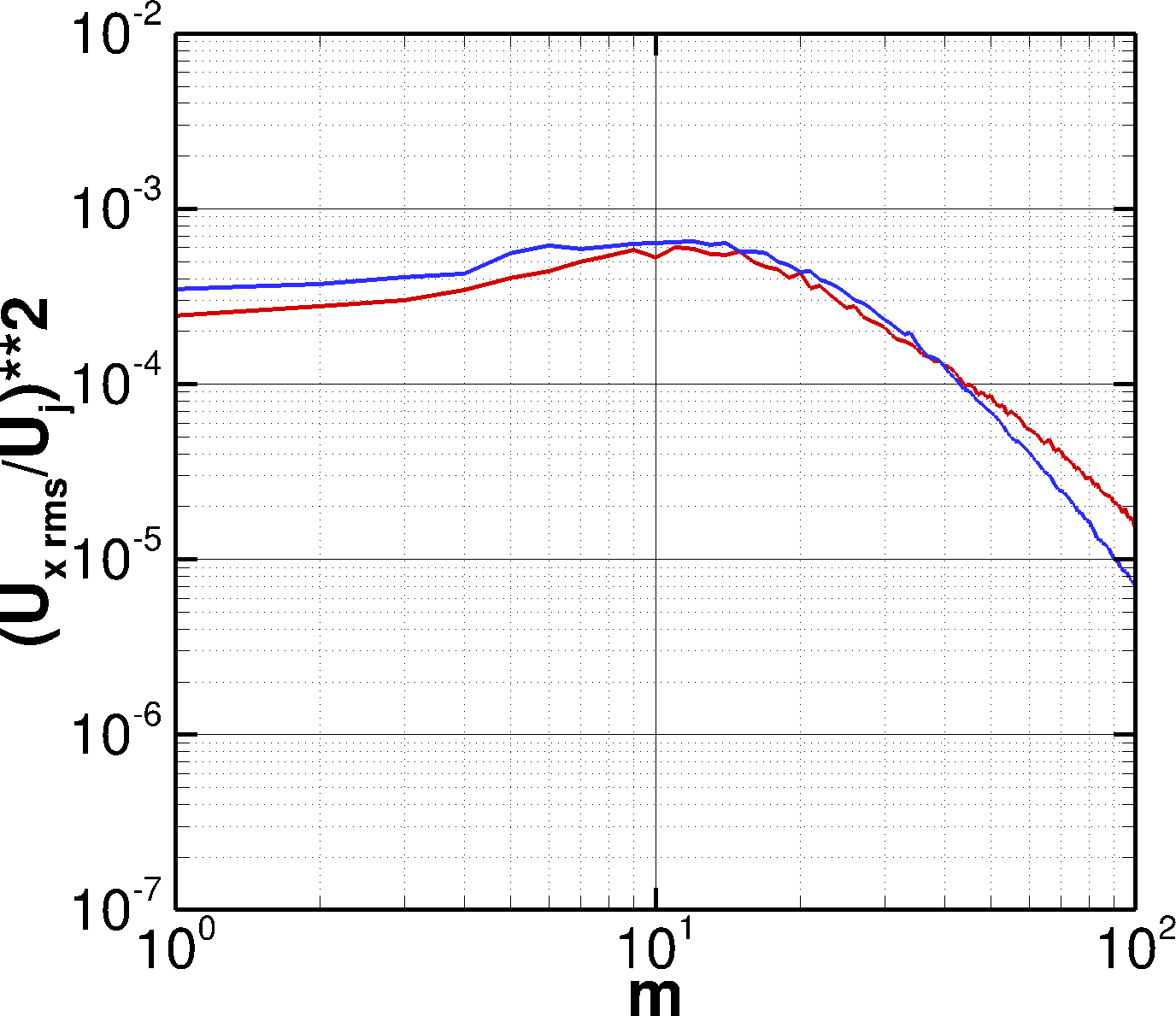}
		\end{minipage}
		\hfill
		\vspace{0.1cm}
		\begin{minipage}[c]{0.32\linewidth}
			\centering (a) $x/D_j = 0.25$
		\end{minipage}
		\hfill
		\vspace{0.1cm}
		\begin{minipage}[c]{0.32\linewidth}
			\centering (b) $x/D_j = 0.5$
		\end{minipage}
		\hfill
		\vspace{0.1cm}
		\begin{minipage}[c]{0.32\linewidth}
			\centering (c) $x/D_j = 1$
		\end{minipage}
		\hfill
		\begin{minipage}[c]{\linewidth}
			\vspace{0.1cm}
	 		\caption[]{Comparison of rms axial fluctuating velocity adimensioned by $U_j$ at $r/D_j = 0.5$ for $x/D_j \in \{0.25; 0.5; 1\}$ as function of the azimuthal mode order $m$. \textcolor{red}{\trait}, \emph{Aghora} simulation; \textcolor{blue}{\trait}, CEDRE simulation}
 			\label{fig_ModesAzim_Ux_cisail_COMP_aghora_cedre}
		\end{minipage}	
	\end{center}
\vspace{-0.8cm}
\end{figure}

The azimuthal composition of the fluctuating axial velocity field is shown in figure~\ref{fig_ModesAzim_Ux_cisail_COMP_aghora_cedre}, for both simulations at the same axial positions as in figure~\ref{fig_DSP_Ux_cisail_COMP_aghora_cedre}. The two simulations have different azimuthal compositions for the positions close to the nozzle exit ($x/D_j \leq 0.5$). At $x/D_j = 0.25$, the \emph{Aghora} simulation has higher rms levels than the CEDRE solution (see figure~\ref{fig_Ux_rms_max_aghora_cedre}) but the velocity field is dominated by the same azimuthal modes $|m| \leq 8$. At $x/D_j = 0.5$ both simulations tend to have similar azimuthal compositions. However, we observe that the \emph{Aghora} simulation presents a higher energy content at the upper part of the spectrum, compared to the FV results.
 For axial positions $x/D_j \geq 1$, both simulations have similar azimuthal compositions and we can see that as we move downstream the exit the azimuthal structure is dominated by lower and lower azimuthal modes.
\\

We have seen from these analysis that the two simulations exhibit differences for the shear layer close to the nozzle exit ($x/D_j \leq 0.5$), whereas for downstream positions both solutions behave similarly. We will now study how these differences impact the jet development.

\subsection{Jet development}
\label{subsec_JetDev}

\begin{figure}[h!]
	\begin{center}
		\begin{minipage}[c]{0.49\linewidth}
			\centering \includegraphics[height=4cm]{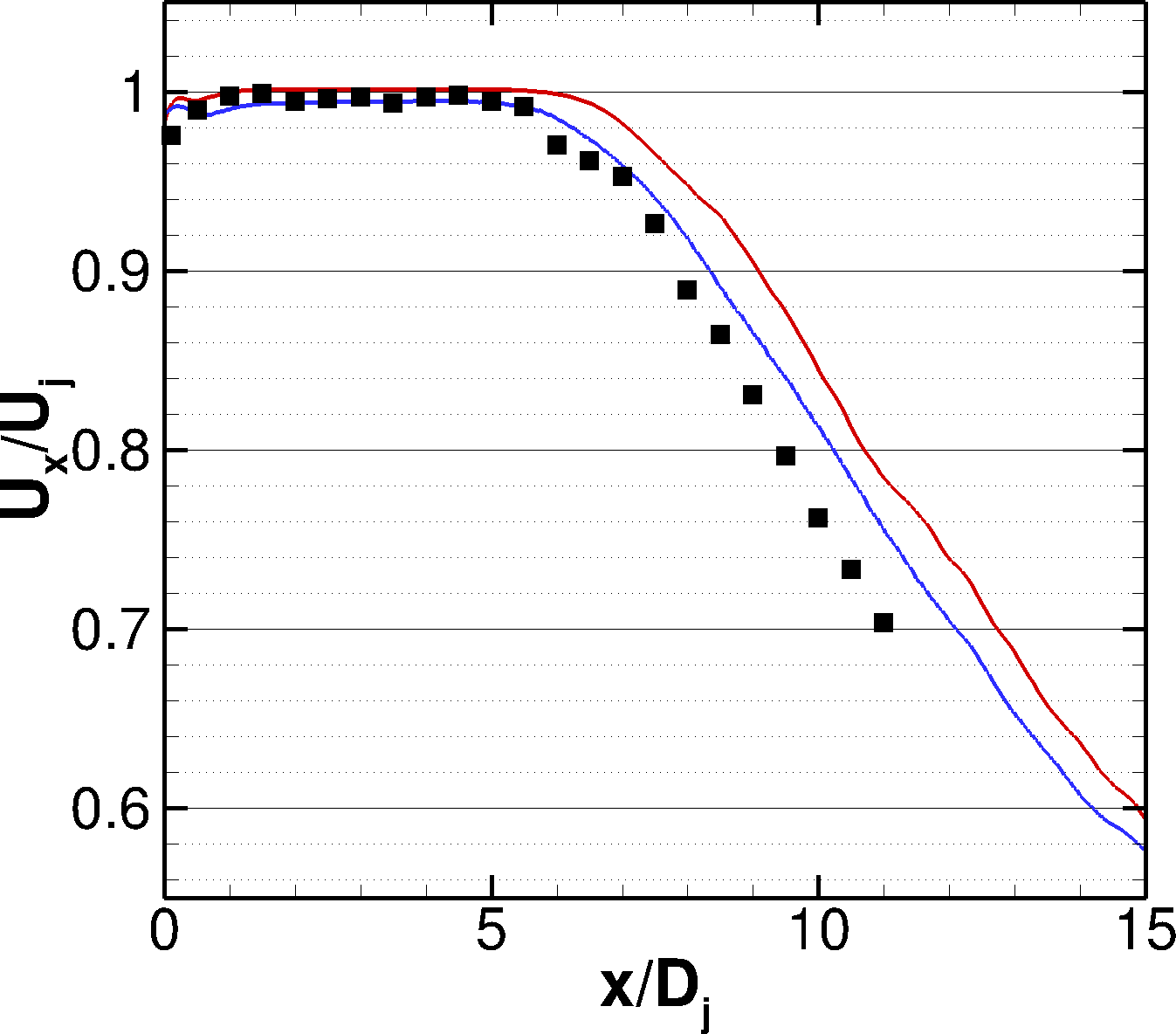}
		\end{minipage}
		\hfill
		\begin{minipage}[c]{0.49\linewidth}
			\centering \includegraphics[height=4cm]{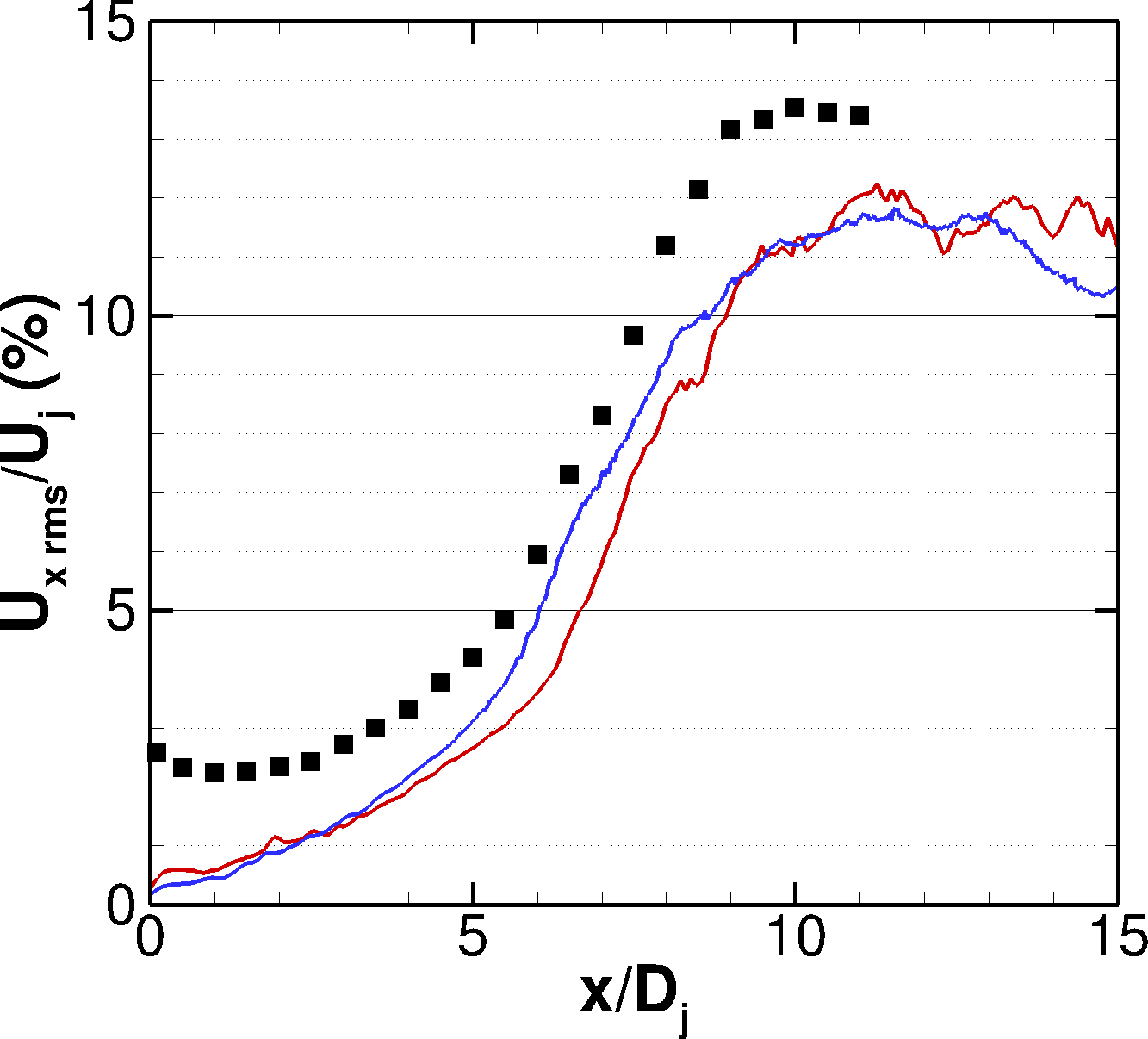}
		\end{minipage}
		\hfill
		\vspace{0.1cm}
		\begin{minipage}[c]{0.49\linewidth}
			\centering (a)
		\end{minipage}
		\hfill
		\vspace{0.1cm}
		\begin{minipage}[c]{0.49\linewidth}
			\centering (b)
		\end{minipage}
		\hfill
		\begin{minipage}[c]{\linewidth}
			\vspace{0.1cm}
	 		\caption[]{Comparison of the axial evolution on the jet axis of the (a) mean and (b) rms axial velocity. \textcolor{red}{\trait}, \emph{Aghora} simulation; \textcolor{blue}{\trait}, CEDRE simulation;  $\blacksquare$, measurements}
 			\label{fig_Ux_moy_rms_axe_COMP_aghora_cedre}
		\end{minipage}	
	\end{center}
\vspace{-0.8cm}
\end{figure}

\hspace{0.5cm} The shear layer analysed in the previous section has an impact on the jet development. As can be seen in figure~\ref{fig_Ux_moy_rms_axe_COMP_aghora_cedre}, the two simulations have a longer potential core length $L_c$ (with $L_c$ taken as $U_x(L_c) = 0.95 \times U_j$)  than that found in the measurements. The \emph{Aghora} simulation yields $L_c/D_j = 7.9$ which is longer than in the CEDRE simulation for which $L_c/D_j = 7.3$, whereas the measured value is $L_c/D_j = 7$. 

We can also observe from figure~\ref{fig_Ux_moy_rms_axe_COMP_aghora_cedre}(b) that despite having an axial location of the peak of rms axial velocity in agreement with the experimental data, the rms levels are lower in the simulation ($\sim 12\%$ against $13.5\%$). This might be an effect of the initial levels of rms velocity on the jet axis which are lower in the simulations than those obtained experimentally. 

The increase of the length of the potential core and the lower rms velocity levels on the jet axis is not in agreement with what is usually observed in simulated jets of this type. Indeed, a laminar shear layer present at the nozzle exit leads to a shorter potential core and higher 
 rms velocity levels on the jet axis (see \cite{Bogey2012,Lorteau2015} for instance). Another effect that might explain this behaviour has been highlighted by Bogey and Marsden \cite{Bogey2010} who have shown that increasing the boundary layer thickness at constant jet diameter causes a quicker jet spreading starting further downstream. Here, these two, or more, effects might be the cause of the increase in the length of the potential core and the lower levels of rms velocity (in accordance with a longer potential core) found along the jet axis. We will now see how this has an impact on the pressure in the far field.

\subsection{Far-field pressure}
\label{subsec_FarField}

\begin{figure}[h!]
	\begin{center}
		\begin{minipage}[c]{\linewidth}
			\centering \includegraphics[height=4cm]{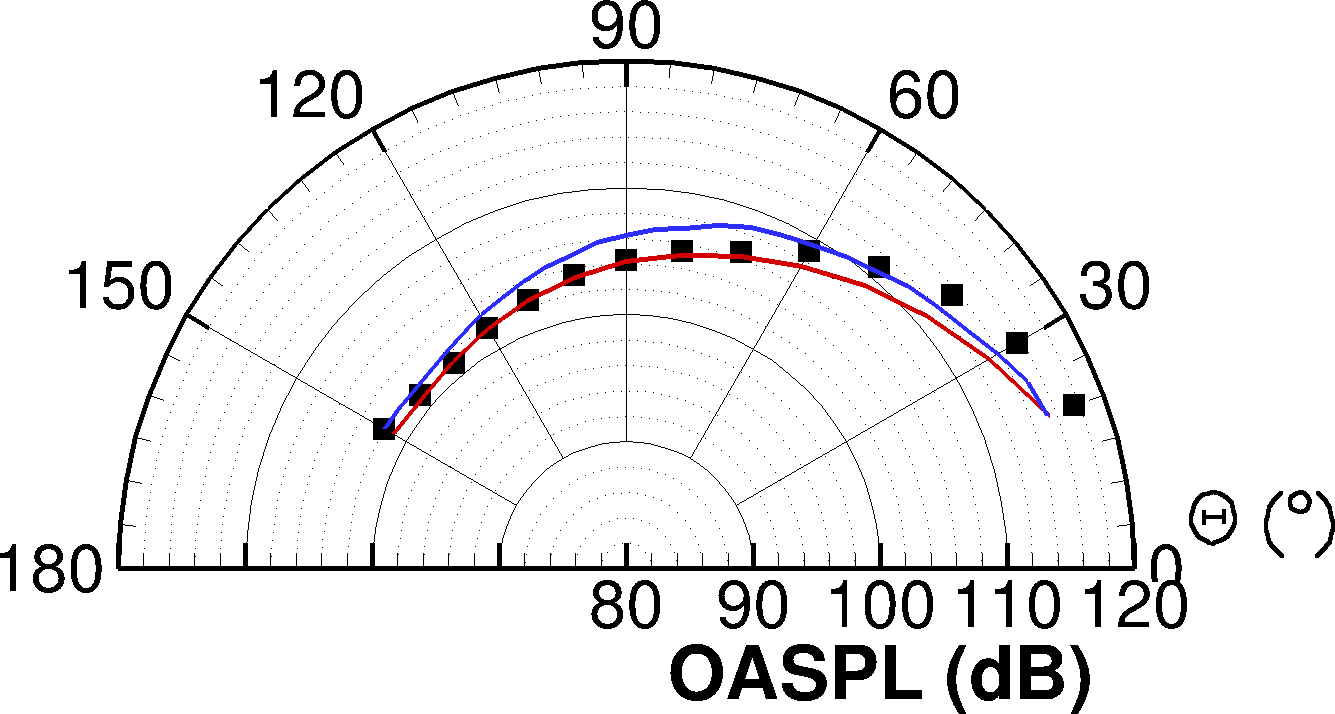}
		\end{minipage}
		\hfill
		\begin{minipage}[c]{\linewidth}
			\vspace{0.1cm}
	 		\caption[]{OASPL of the pressure field in the far field at $30D_j$ from the nozzle exit. \textcolor{red}{\trait}, \emph{Aghora} simulation; \textcolor{blue}{\trait}, CEDRE simulation; $\blacksquare$, measurements}
 			\label{fig_OASPL_loin_30D_COMP_aghora_cedre}
		\end{minipage}	
	\end{center}
\vspace{-0.8cm}
\end{figure}

\hspace{0.5cm} In this work, the angular position $30^\circ$ represents the downstream direction whereas $150^\circ$ represents the upstream direction. The angular position $90^\circ$ corresponds to the jet exit plane ($x/D_j = 0$). We can see in figure~\ref{fig_OASPL_loin_30D_COMP_aghora_cedre} that the overall pressure level (OASPL) directivity in the far field at $30D_j$ from the nozzle exit is well predicted within a $\pm 2$~dB margin. Detailed analysis shows that there is a slight underestimation of the measured level for $\theta \in [20^\circ; 60^\circ]$ for both simulations and a slight overestimation for $\theta \in [60^\circ; 120^\circ]$ in the CEDRE simulation. For higher angles, i.e.\ $\theta \geq 120^\circ$, a good agreement is observed for both simulations.
\\

\begin{figure}[h!]
	\begin{center}
		\begin{minipage}[c]{0.32\linewidth}
			\centering \includegraphics[height=4cm]{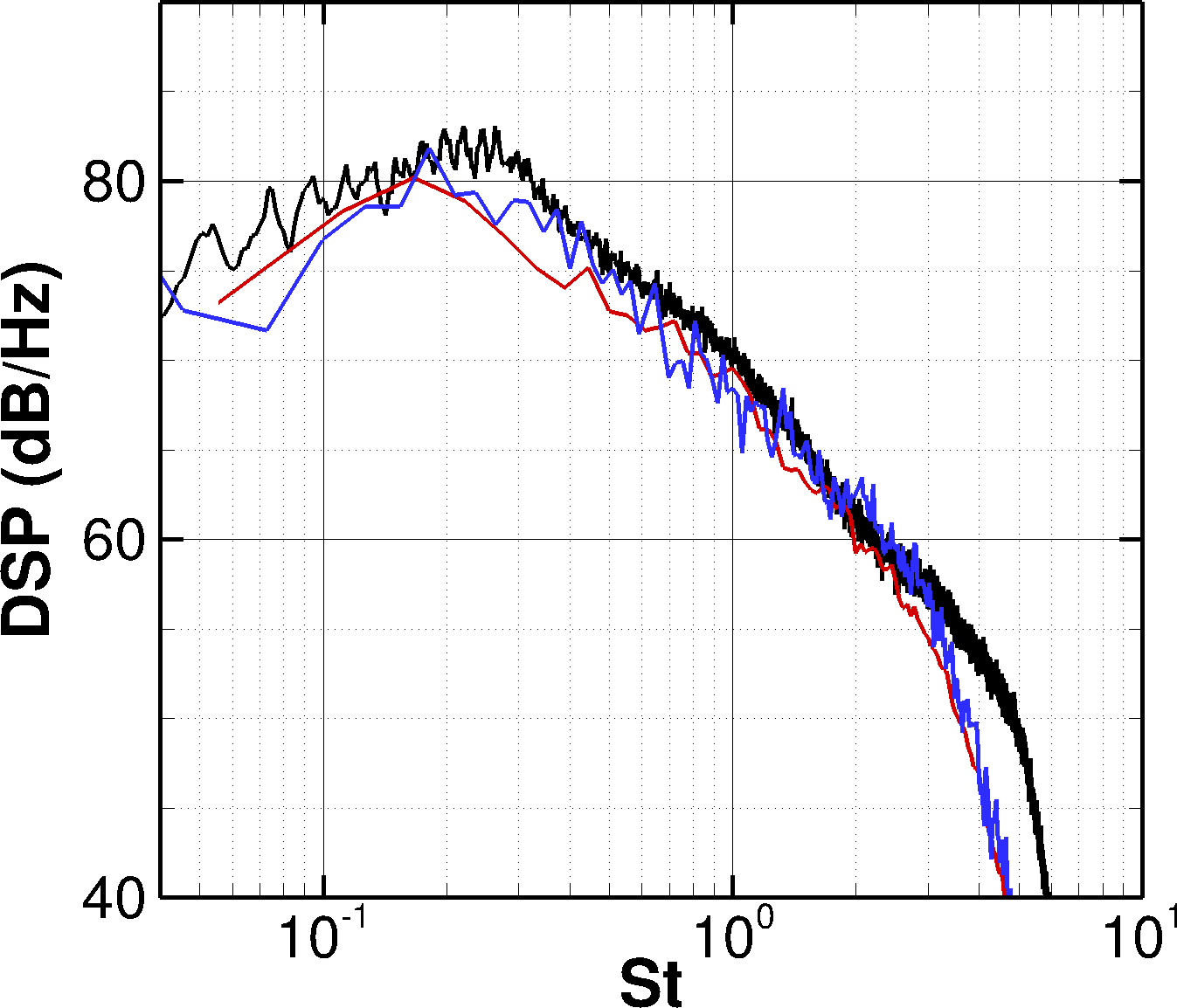}
		\end{minipage}
		\hfill
		\begin{minipage}[c]{0.32\linewidth}
			\centering \includegraphics[height=4cm]{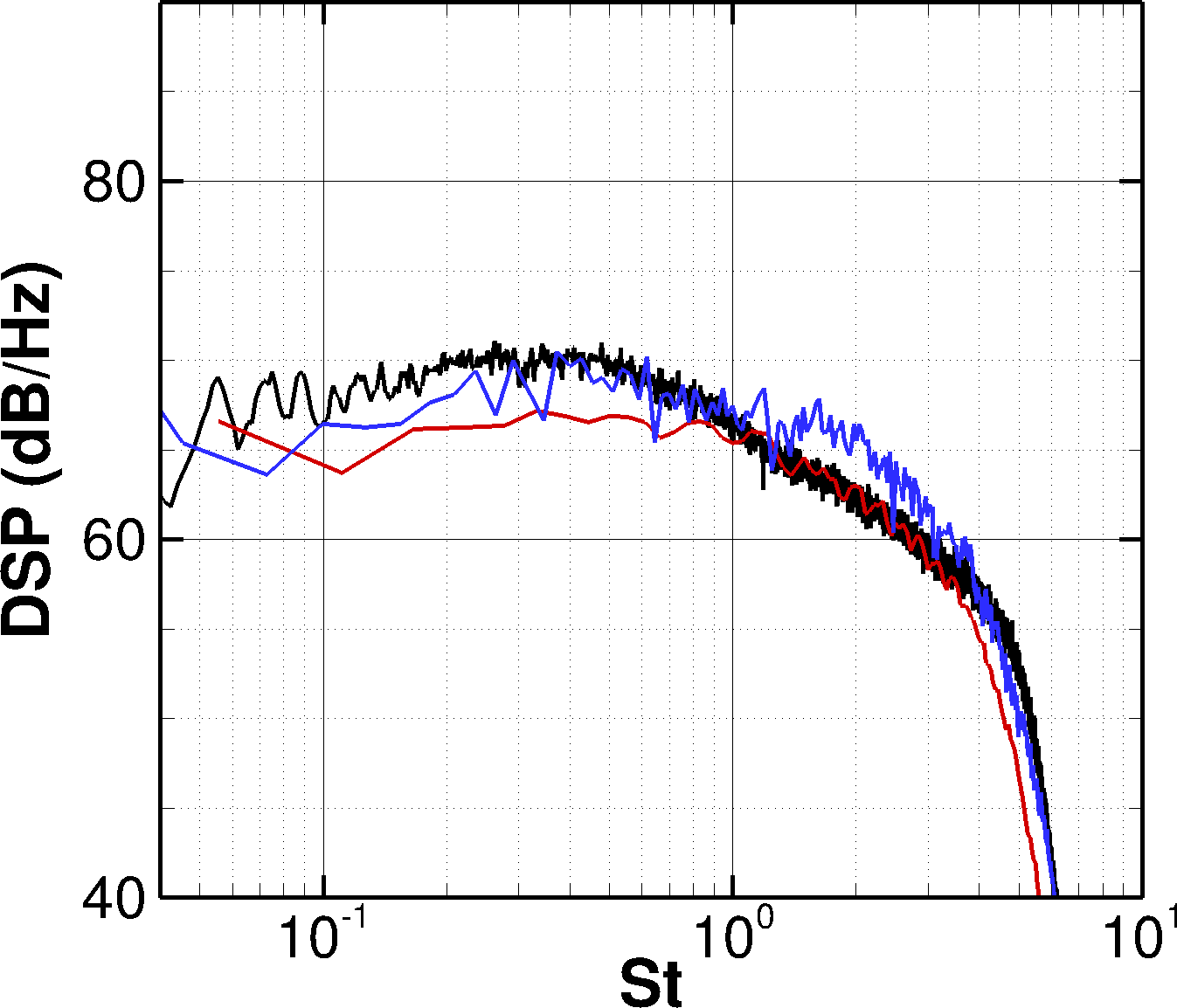}
		\end{minipage}
		\hfill
		\begin{minipage}[c]{0.32\linewidth}
			\vspace{0.1cm}
			\centering \includegraphics[height=4cm]{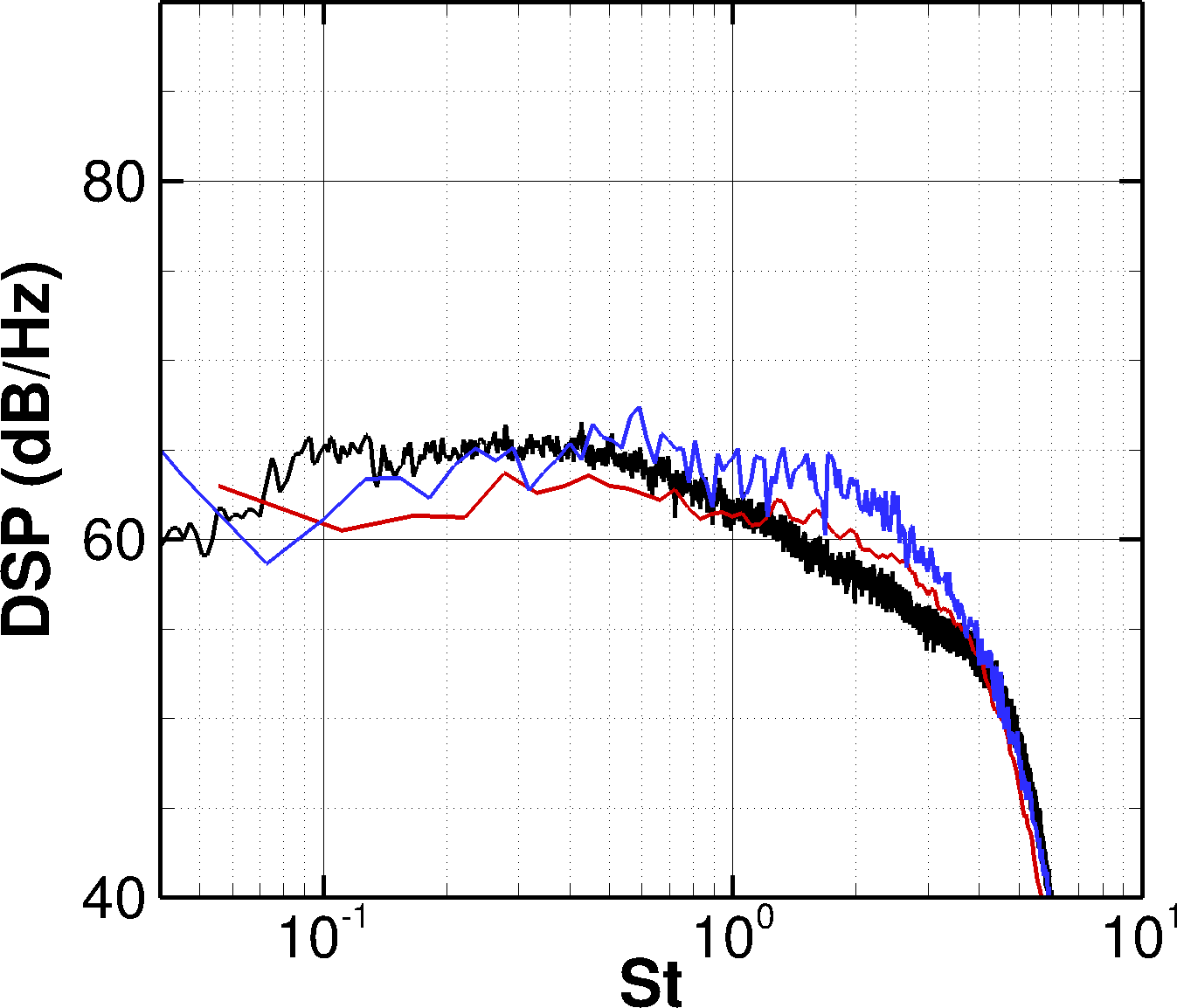}
		\end{minipage}
		\hfill
		\vspace{0.1cm}
		\begin{minipage}[c]{0.32\linewidth}
			\vspace{0.1cm}
			\centering (a) $30^\circ$
		\end{minipage}
		\hfill
		\vspace{0.1cm}
		\begin{minipage}[c]{0.32\linewidth}
			\vspace{0.1cm}
			\centering (b) $60^\circ$
		\end{minipage}
		\hfill
		\vspace{0.1cm}
		\begin{minipage}[c]{0.32\linewidth}
			\vspace{0.1cm}
			\centering (c) $90^\circ$
		\end{minipage}
		\hfill
		\begin{minipage}[c]{\linewidth}
			\vspace{0.1cm}
	 		\caption[]{PSDs of the pressure field in the far field at $30D_j$ from the nozzle exit for different angular positions. \textcolor{red}{\trait}, \emph{Aghora} simulation; \textcolor{blue}{\trait}, CEDRE simulation; \textcolor{black}{\trait}, measurements}
 			\label{fig_DSP_loin_30D_COMP_aghora_cedre}
		\end{minipage}	
	\end{center}
\vspace{-0.8cm}
\end{figure}

The PSDs presented in figure~\ref{fig_DSP_loin_30D_COMP_aghora_cedre} allow us to interpret these observations. We can see that for all the angular positions, there is an underestimation of the low frequency part (i.e.\ $St \leq 0.5$) of the spectra for both simulations. This is more pronounced in the \emph{Aghora} simulation. This underestimation of the low frequency content leads to an underestimation of the integrated levels in the downstream angular sector where they dominate the acoustic radiation. Moreover it has been shown that these low frequencies dominating the acoustic radiation are linked to what happens around the potential core end (see Tam \emph{et al.}\cite{Tam2008}). Thus, the underestimation of the low frequency content may be associated to the underestimation of the rms velocity levels on the jet axis (cf. figure~\ref{fig_Ux_moy_rms_axe_COMP_aghora_cedre}(b)) or in the shear layer (cf. figure~\ref{fig_Ux_rms_max_aghora_cedre}).

When we move from the downstream direction to the upstream direction, we can see that there is an overestimation of the high frequencies (for $St \in [1; 3]$). This is even more pronounced in the CEDRE simulation. This overestimation seems to balance the underestimation in the low frequencies. This can explain the good agreement found with the measured integrated levels (even a slight overestimation for the CEDRE simulation). The overestimation of the high-frequency levels found in the simulations can be linked 
to the transition of the shear layer from laminar to turbulent. Before turbulence is fully developed, pairings occur leading to an additional acoustic source. This source is especially visible for an angular sector around $90^\circ$ where pressure levels are lower \cite{Bogey2012,Zaman1985_1}. This effect is less pronounced for the \emph{Aghora} simulation despite its shear layer being laminar at the nozzle exit. This can be related to the thinner shear layer of the \emph{Aghora} simulation, as it was shown by Bogey \emph{et al.}~\cite{Bogey2010} and Zaman~\cite{Zaman1985_1} that a thinner shear layer gives a lower overestimation due to pairings.
\\

In spite of having the same jet configuration and similar meshes, the two simulations display some differences in the shear layer state which influence the flow field and therefore the radiated acoustic field. These differences might be linked to the different properties presented by their respective numerical schemes.

\section{Improvement of the DG simulation}
\label{sec_HexaP3}


\hspace{0.5cm} The DG simulation presented in the previous section has been performed on a tetrahedral mesh and we have seen that the simulated jet does not reproduce accurately the experimental data, especially in the shear layer close to the nozzle exit. This section is devoted to possible ways of accuracy improvements thanks to the flexible capacities of the DG approach. We will here study the influence of the mesh, and its associated set of basis functions, through detailed analysis on two fundamental test cases.

\subsection{Analysis of the DG method on fundamental test cases}
\label{subsec_PropSchema}

\hspace{0.5cm} In this section, we assess the numerical method used. To this end, we consider two different configurations: the Taylor-Green vortex at $Re = 1600$ \cite{Taylor1937} and the propagation of an oblique acoustic wave. For these two cases, the computations are run on a number of different meshes: regular hexahedra and tetrahedra and at different resolutions. Different polynomial orders are also used. These two test cases will allow us to study the influence of the basis functions as a given type of element is associated to a set of basis functions, for instance a tetrahedron is associated to a Dubiner basis whereas a hexahedron is associated to a tensored basis.

\subsubsection{Taylor Green Vortex analysis}
\label{subsubsec_TGV}

\hspace{0.5cm} The Taylor-Green vortex configuration represents a transitional freely decaying turbulent flow in a periodic box. A velocity profile representing large vortical structures is applied as the initial conditions. These vortices break up into smaller structures as time evolves generating an energy cascade. The configuration under consideration corresponds to $Re = 1600$ with $M = 0.1$.

Three different resolutions (i.e.\ number of DoFs) have been tested: 64$^3$, 128$^3$ and 256$^3$. Two types of meshes are considered: regular hexahedral cells and tetrahedral cells. For the hexahedral meshes, polynomial degrees $2 \leq p \leq 5$ have been used, whereas for the tetrahedral meshes only polynomial degrees $p = 2$ and $p = 5$ are considered.

The time integration, convective and viscous fluxes discretization methods are the same as those used to simulate the configuration and described in section~\ref{subsec_NumMethod}.

\begin{figure}[h!]
	\begin{center}
		\begin{minipage}[c]{0.32\linewidth}
			\centering \includegraphics[height=3.5cm]{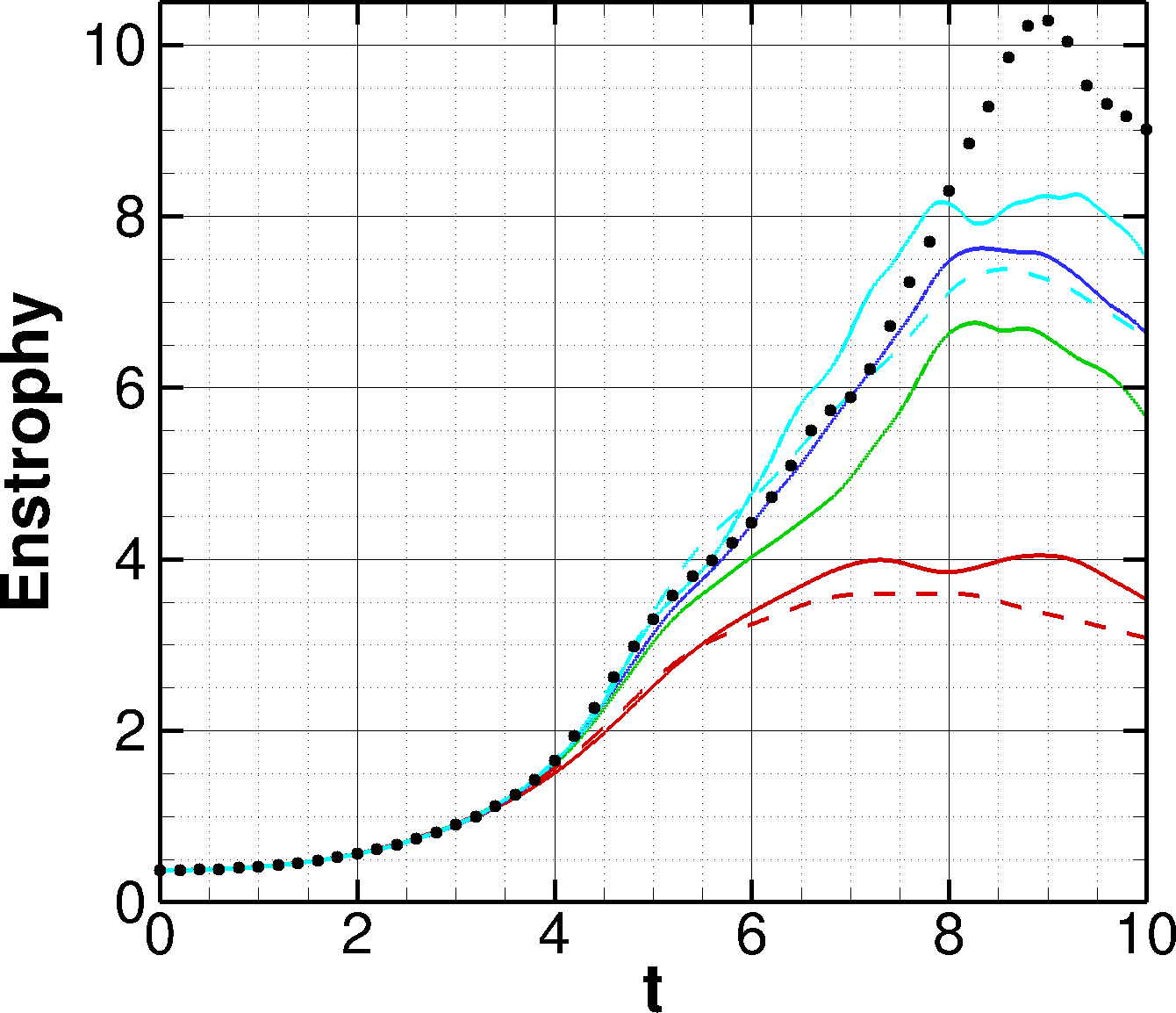}
		\end{minipage}
		\hfill
		\begin{minipage}[c]{0.32\linewidth}
			\centering \includegraphics[height=3.5cm]{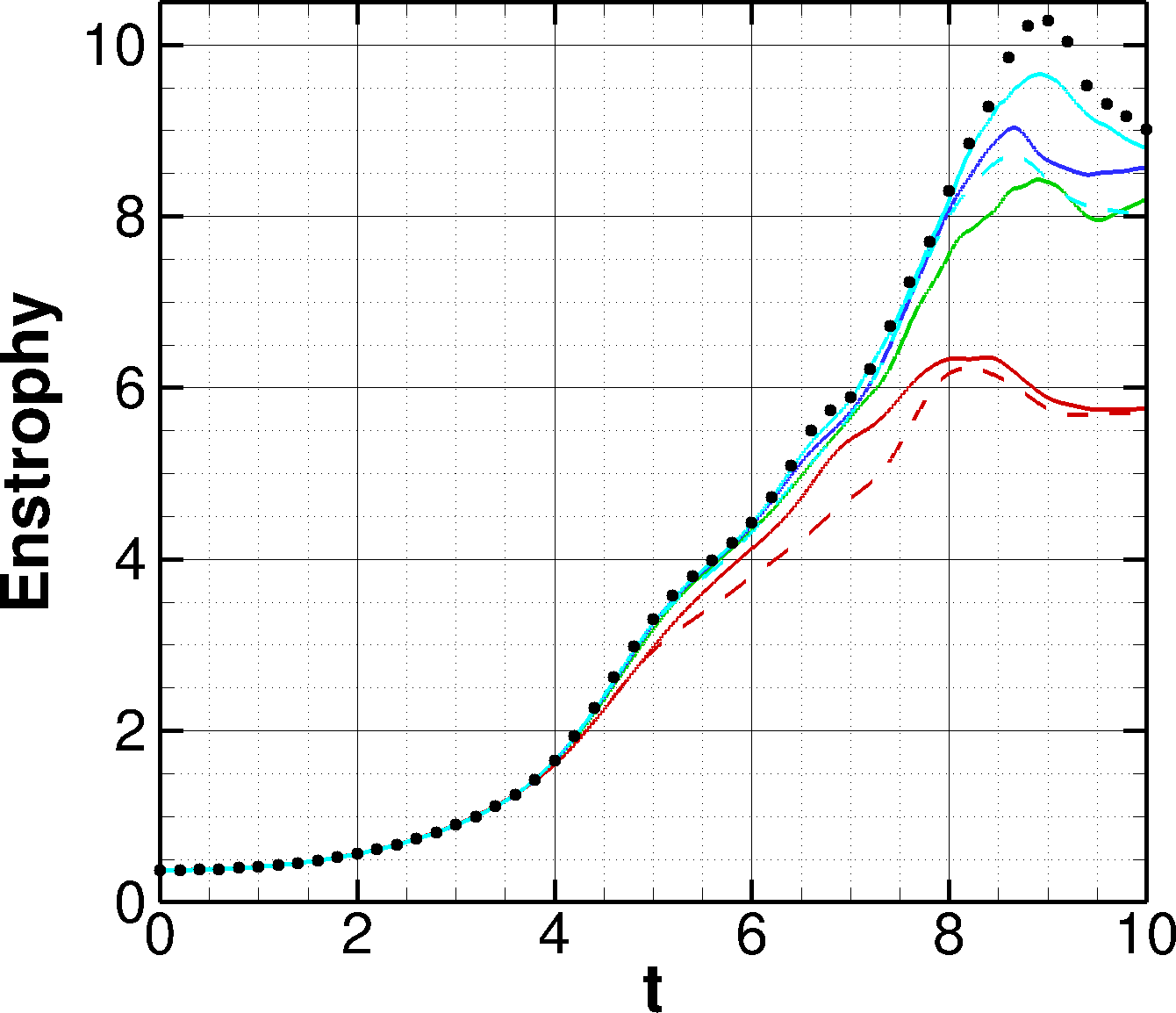}
		\end{minipage}
		\hfill
		\begin{minipage}[c]{0.32\linewidth}
			\centering \includegraphics[height=3.5cm]{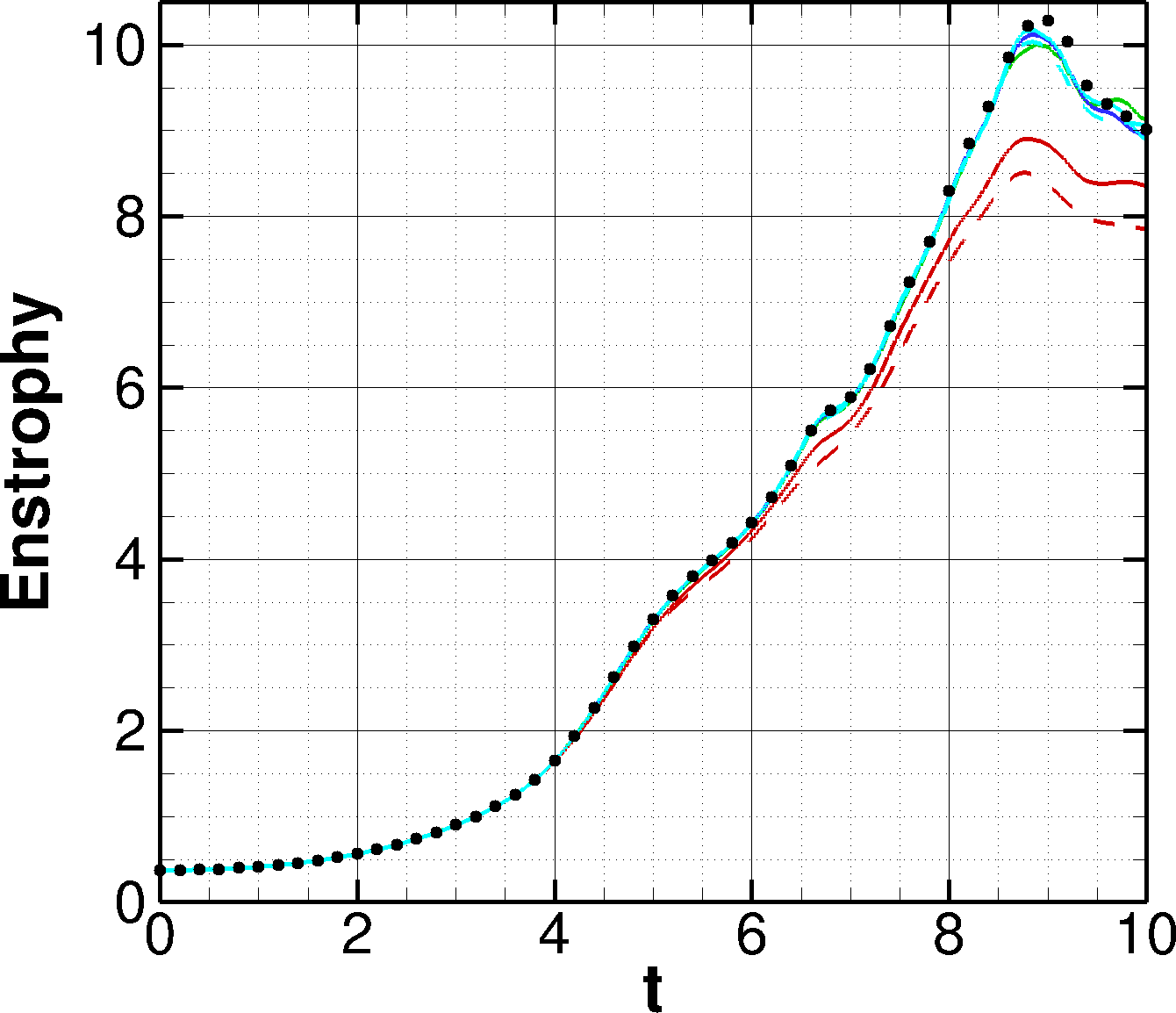}
		\end{minipage}
		\hfill
		\vspace{0.1cm}
		\begin{minipage}[c]{0.32\linewidth}
			\centering (a) 64$^3$ DoFs
		\end{minipage}
		\hfill
		\vspace{0.1cm}
		\begin{minipage}[c]{0.32\linewidth}
			\centering (b) 128$^3$ DoFs
		\end{minipage}
		\hfill
		\vspace{0.1cm}
		\begin{minipage}[c]{0.32\linewidth}
			\centering (b) 256$^3$ DoFs
		\end{minipage}
		\hfill
		\begin{minipage}[c]{\linewidth}
	 		\caption[]{Comparison of the temporal evolution of the enstrophy for the three resolutions for different polynomial degrees. \textcolor{red}{\trait}, $p = 2$; \textcolor{green}{\trait}, $p = 3$; \textcolor{blue}{\trait}, $p = 4$; \textcolor{cyan}{\trait}, $p = 5$; \trait, hexahedral mesh; \tirets, tetrahedral mesh; $\bullet$, reference DNS}
 			\label{fig_TGV_enstrophy_64_128_256}
		\end{minipage}	
	\end{center}
\vspace{-0.5cm}
\end{figure}

Figure~\ref{fig_TGV_enstrophy_64_128_256} depicts the evolution of the enstrophy for all computations performed. A DNS with a 512$^3$ resolution performed using a Fourier spectral code is also shown as a reference, see \cite{Chapelier2014}. The enstrophy is the physical value presented here. The other turbulent quantities (e.g.\ kinetic energy dissipation) were not as discriminative as this one. As expected, the finer the resolution, the better the results. Moreover it can be seen that, even at a similar resolution and polynomial degree, a simulation performed on a hexahedral mesh yields more accurate results than one performed on a tetrahedral mesh. We can also see that even with less DoFs, a $p = 5$ simulation can achieve a finer solution than a $p = 2$ simulation, e.g.\ $p = 5$ with 128$^3$ DoFs as compared to $p = 2$ with 256$^3$ DoFs. This observation was already pointed out in \cite{Chapelier2014}.

\begin{figure}[h!]
	\begin{center}
		\begin{minipage}[c]{\linewidth}
			\centering \includegraphics[height=5cm]{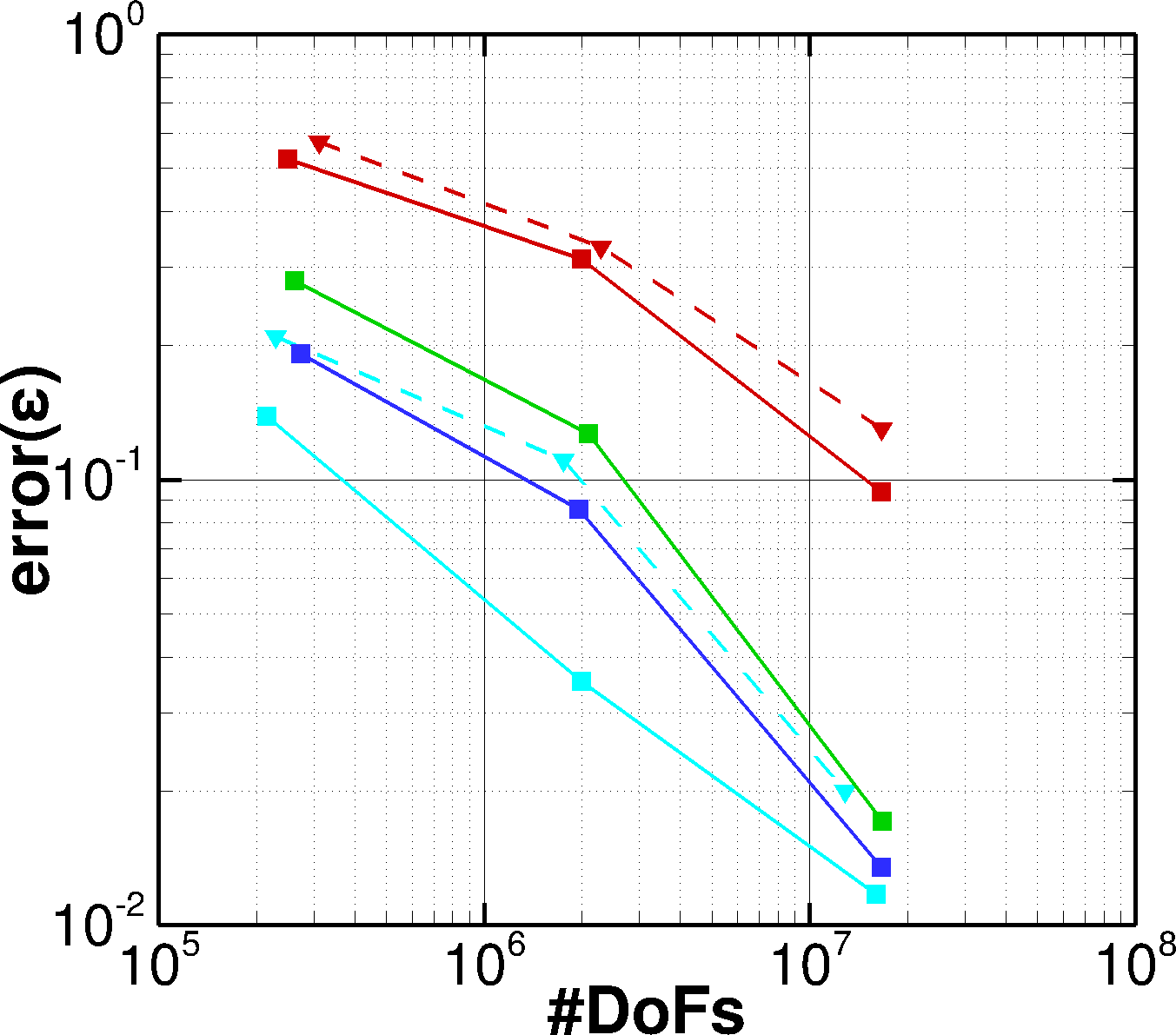}
		\end{minipage}
		\hfill
		\begin{minipage}[c]{\linewidth}
			\vspace{0.1cm}
	 		\caption[]{Comparison of the error on the enstrophy $\epsilon$ as a function of the number of the number of DoF. \textcolor{red}{\trait}, $p = 2$; \textcolor{green}{\trait}, $p = 3$; \textcolor{blue}{\trait}, $p = 4$; \textcolor{cyan}{\trait}, $p = 5$; (\trait, $\blacksquare$), hexahedral mesh; (\tirets, $\blacktriangledown$), tetrahedral mesh}
 			\label{fig_TGV_erreur_NDoFs}
		\end{minipage}	
	\end{center}
\vspace{-0.5cm}
\end{figure}

These latter results can be better quantified in figure~\ref{fig_TGV_erreur_NDoFs} representing the error on the enstrophy as a function of the number of DoFs for all computations performed. The error is computed using the same reference DNS presented in figure~\ref{fig_TGV_enstrophy_64_128_256}. From these results, it can be concluded that a polynomial degree $p = 3$ is at least required to achieve a sufficiently fine solution (for the resolutions under consideration).

\subsubsection{Acoustic dissipation of a plane oblique wave}
\label{subsubsec_CAA}

\hspace{0.5cm} Here, we study the upstream propagation of an oblique acoustic wave. More precisely we want to evaluate the number of cells per wavelength (\emph{cpw}) at a fixed polynomial degree necessary to correctly propagate an acoustic wave through a given computational domain.

\begin{figure}[h!]
	\begin{center}
		\begin{minipage}[c]{0.49\linewidth}
			\centering \includegraphics[height=1.75cm]{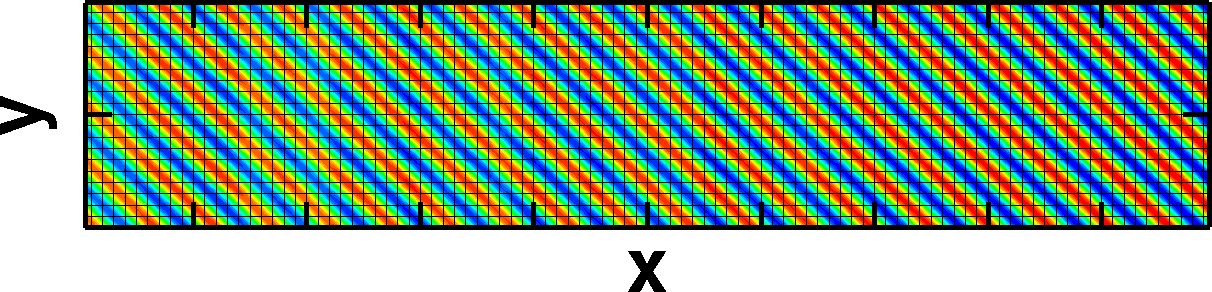}
		\end{minipage}
		\hfill
		\begin{minipage}[c]{0.49\linewidth}
			\centering \includegraphics[height=1.75cm]{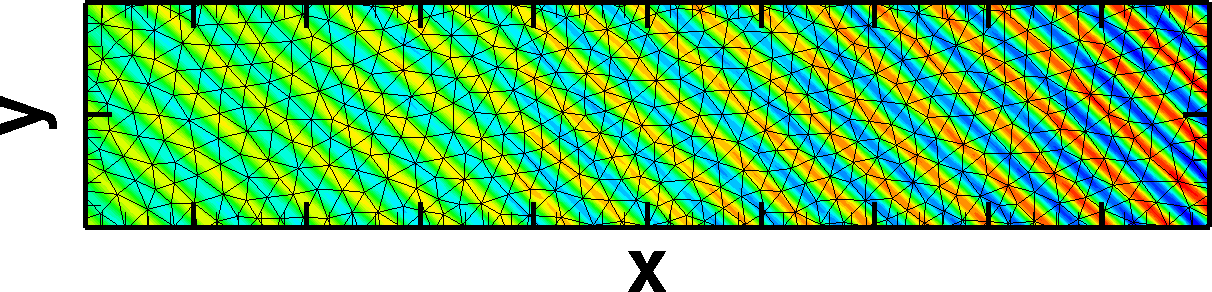}
		\end{minipage}
		\hfill
		\begin{minipage}[c]{0.49\linewidth}
			\vspace{0.1cm}
			\centering (a) quadrangular mesh
		\end{minipage}
		\hfill
		\begin{minipage}[c]{0.49\linewidth}
			\vspace{0.1cm}
			\centering (b) triangular mesh
		\end{minipage}
		\hfill
		\begin{minipage}[c]{\linewidth}
			\vspace{0.1cm}
	 		\caption{View of the propagation of an acoustic wave for both meshes with $p = 3$ and 5 \emph{cpw}.}
 			\label{fig_CAA_hexa_TetraUns_p3_5ppw}
		\end{minipage}	
	\end{center}
\vspace{-0.5cm}
\end{figure}

A backward travelling oblique acoustic wave is prescribed through a time dependent static pressure at the outlet at an amplitude of 2~Pa and in flow at Mach number $M = 0.4$. A multidimensional non-reflecting boundary condition \cite{Couaillier2005} is imposed at the inlet.
As for the Taylor-Green vortex case, two types of meshes are tested: regular hexahedral cells and tetrahedral cells. The meshes are illustrated in figure~\ref{fig_CAA_hexa_TetraUns_p3_5ppw}. For both meshes, polynomial degrees $1 \leq p \leq 5$ and $\{1; 2; 5; 10\}$ \emph{cpw} have been considered. In practice, the mesh is fixed and the wavelength is modified, so that the mesh is defined for the largest wavelength under study in order to propagate at least 10 wavelengths. This leads to a computational domain of $L_x = 0.1$ and $L_y = 0.02$.

The computations have been performed in 2D. It has been verified that, in this case, the results are similar to those obtained on a 3D mesh. Thus, for this study we are using a regular quadrangular mesh and a triangular mesh rather than a regular hexahedral mesh and a tetrahedral mesh respectively. If we consider that in a jet flow there are almost practically no viscous effect in the outer region, that is to say the region in which the acoustic waves propagate, we can perform this study by using the Euler equations instead of the Navier-Stokes equations. The explicit third-order accurate Runge-Kutta method and the local Lax-Friedrichs (LLF) flux previously mentioned have been used for the computations considered in this section.

In figure~\ref{fig_CAA_hexa_TetraUns_p3_5ppw}, the acoustic wave propagation is represented for the two meshes for $p = 3$ and 5 \emph{cpw}. The dissipation properties of the two discretizations are clearly highlighted in this figure.

To better quantify the quality of the discretization (mesh and p) for each wavelength, we extract the pressure values along the middle line of the computational domain (i.e.\ $y = L_y/2$) and fit this longitudinal evolution with a pressure wave of the form:

\vspace{-0.2cm}
\begin{equation}
 p(x) = p_0e^{-\alpha x}sin(2\pi x/\lambda+\phi)
\label{equ_CAA_fit}
\end{equation}

\noindent where $\alpha$ represents the dissipation, $\lambda$ the estimated wavelength and $\phi$ the estimated phase. Thus, with $e^{-\alpha \lambda}$ we can estimate the dissipation of the acoustic wave after the propagation of one wavelength. The dissipation after the propagation of two wavelengths would be $(e^{-\alpha \lambda})^2$. The closer to unity $e^{-\alpha \lambda}$ is, the less dissipated the acoustic wave is. And the dispersion is obtained by comparing the estimated wavelength to the input wavelength. These two parameters are studied below.

\begin{figure}[h!]
	\begin{center}
		\begin{minipage}[c]{0.49\linewidth}
			\centering \includegraphics[height=5cm]{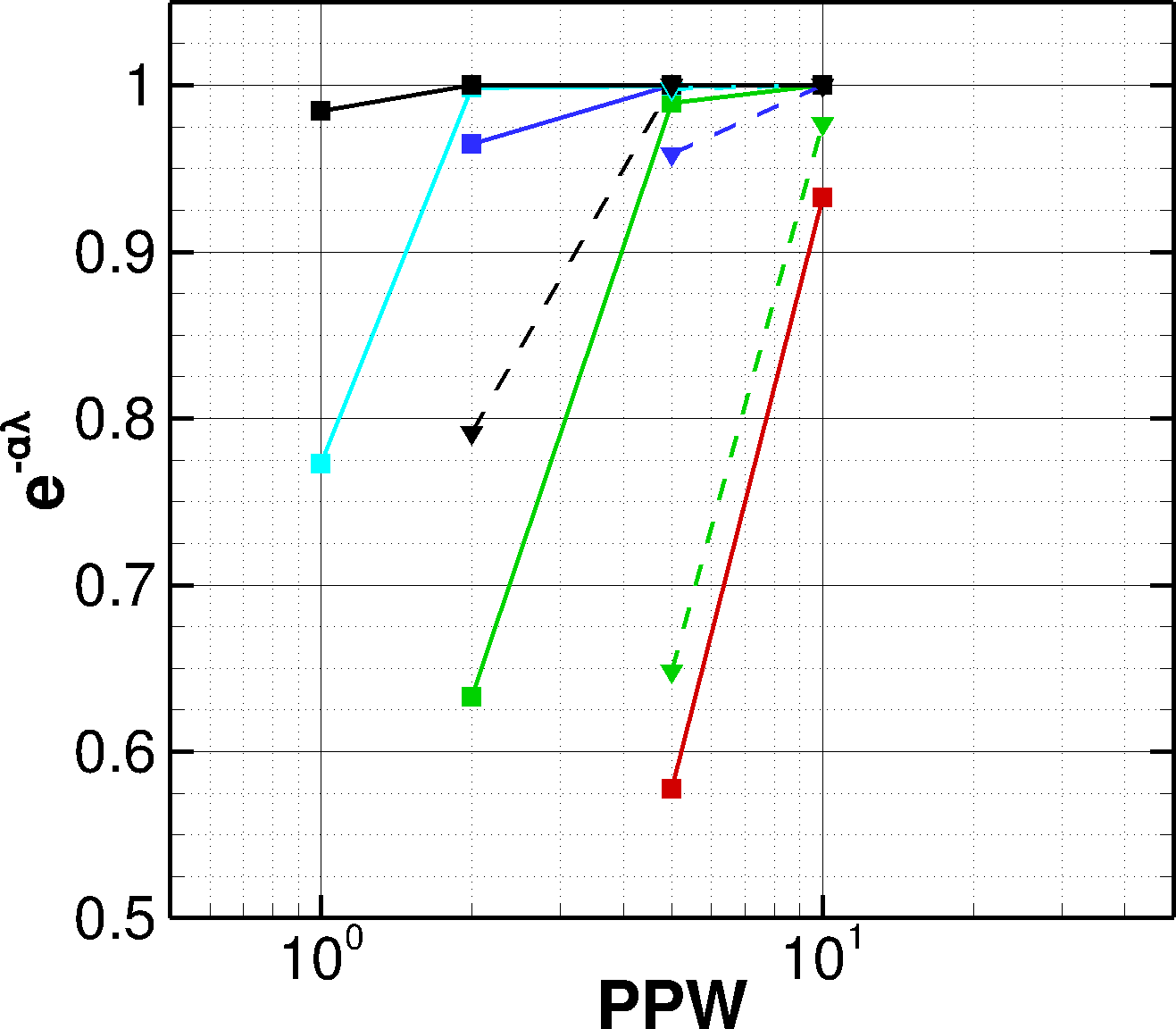}
		\end{minipage}
		\hfill
		\begin{minipage}[c]{0.49\linewidth}
			\centering \includegraphics[height=5cm]{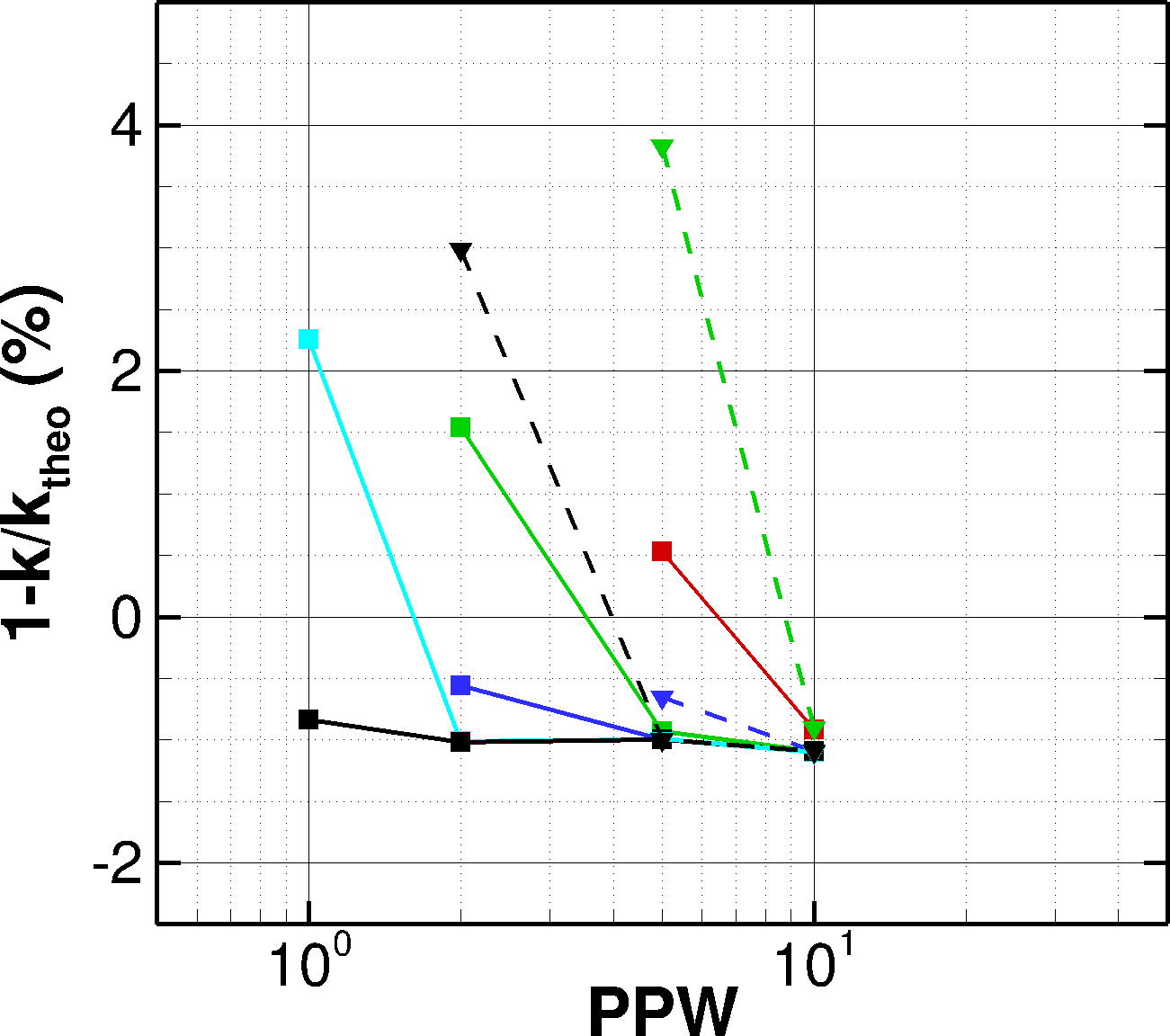}
		\end{minipage}
		\hfill
		\begin{minipage}[c]{0.49\linewidth}
			\vspace{0.1cm}
			\centering (a) dissipation
		\end{minipage}
		\hfill
		\begin{minipage}[c]{0.49\linewidth}
			\vspace{0.1cm}
			\centering (b) dispersion
		\end{minipage}
		\hfill
		\begin{minipage}[c]{\linewidth}
			\vspace{0.1cm}
	 		\caption[]{Comparison of the dissipation (a) and the dispersion (b) of an acoustic as a function of the number of points per wavelength for both meshes and $1 \leq p \leq 5$. \textcolor{red}{\trait}, $p = 1$; \textcolor{green}{\trait}, $p = 2$; \textcolor{blue}{\trait}, $p = 3$; \textcolor{cyan}{\trait}, $p = 4$; \trait, $p = 5$; (\trait, $\blacksquare$), quadrangular mesh; (\tirets, $\blacktriangledown$), triangular mesh}
 			\label{fig_CAA_hexa_TetraUns_dissip_dispers}
		\end{minipage}	
	\end{center}
\vspace{-0.5cm}
\end{figure}

Figure~\ref{fig_CAA_hexa_TetraUns_dissip_dispers} compares the dissipation and dispersion of the injected acoustic wave as a function of the number of cells per wavelength for the quadrangular and the triangular meshes for $1 \leq p \leq 5$. In these pictures, only the cases for which the dissipation is above 0.5 are shown. From these results we can conclude that the number of \emph{cpw} necessary to have a low dissipation (i.e.\ $e^{-\alpha \lambda} \geq 0.95$) and a dispersion less than $1\%$ decreases as the polynomial degree increases. For instance, for the quadrangular mesh, at $p = 2$, 5 \emph{cpw} at least are necessary whereas at $p = 3$, 2 \emph{cpw} are sufficient (even 1 \emph{cpw} for $p = 5$). 

We can also observe that the triangular mesh is more dissipative than the quadrangular one at equal polynomial degrees and number of cells per wavelength. This is expected as for the same polynomial degree $p$, a triangular cell contains less DoFs than a quadrangular cell, namely, $(p+1)(p+2)/2$ compared to $(p+1)^2$ due to the different set of basis functions. It can be argued that the study could have been done at similar resolutions (i.e.\ similar number of DoFs). Nonetheless, it can be seen (see figure~\ref{fig_CAA_hexa_TetraUns_dissip_dispers}(a)) that at least twice (even three times) the number of quadrangles is needed to have an equivalent dissipation for a triangular mesh.

\subsubsection{Synthesis of the basic analysis}
\label{subsubsec_synth}

This study (Taylor-Green vortex and acoustic propagation) has highlighted the benefit of using a hexahedral mesh with a polynomial degree of at least $p \geq 3$ (i.e.\ fourth-order accuracy), associated with a sufficiently fine resolution. The hexahedral meshes allow the use of a full tensored set of basis functions which permits to resolve more frequencies than the Dubiner set of basis functions associated to tetrahedral meshes. Thus, in the following, we are going to use a hexahedral mesh and a polynomial degree $p = 3$ in order to perform the simulation of the jet flow field studied in this work.

\subsection{VMS simulation on a hexahedral mesh}
\label{subsec_HexaP3}

%

\hspace{0.5cm} In this section we present results from the simulation of the studied jet flow on a hexahedral mesh. As observed in section~\ref{sec_Valid}, the numerical dissipation introduced by the Smagorinsky subgrid-scale model appears to play an important role in the jet development. It has thus been decided to limit this dissipation through the use of the \emph{Variational Multiscale Simulation} (VMS) approach~\cite{Chapelier2014}. For this simulation, called HexaP3\_VMS in the following, a value of $C_S = 0.25$ of the Smagorinsky model constant has been used. The presented results are still preliminary, in that the simulated duration is rather short compared to those computed for the simulations previously mentioned (see section~\ref{sec_Valid} and table~\ref{tab_mesh_charac}). Thus the statistics are not fully converged. For the same reason, the far-field pressure is not studied in this section. Nonetheless some interesting tendencies can already be observed. 

\subsubsection{Mesh description}
\label{subsubsec_Mesh_HexaFine}

\hspace{0.5cm} The meshing methodology employed for the hexahedral mesh is the same as the one used for the construction of the tetrahedral mesh. The same computational domain and refined mesh zone as for the tetrahedral mesh have been defined. The maximum mesh size in this zone for the hexahedral mesh is $\Delta x_{max} = 28$~mm. This enables to resolve acoustic waves up to a Strouhal number $St_{cut-off} \approx 1$ with 2 cells per wavelength for the polynomial degree $p = 3$ and the third-order time-marching scheme considered (see section~\ref{subsubsec_CAA}).

\begin{figure}[h!]
	\begin{center}
		\begin{minipage}[c]{\linewidth}
			\centering \includegraphics[height=4cm]{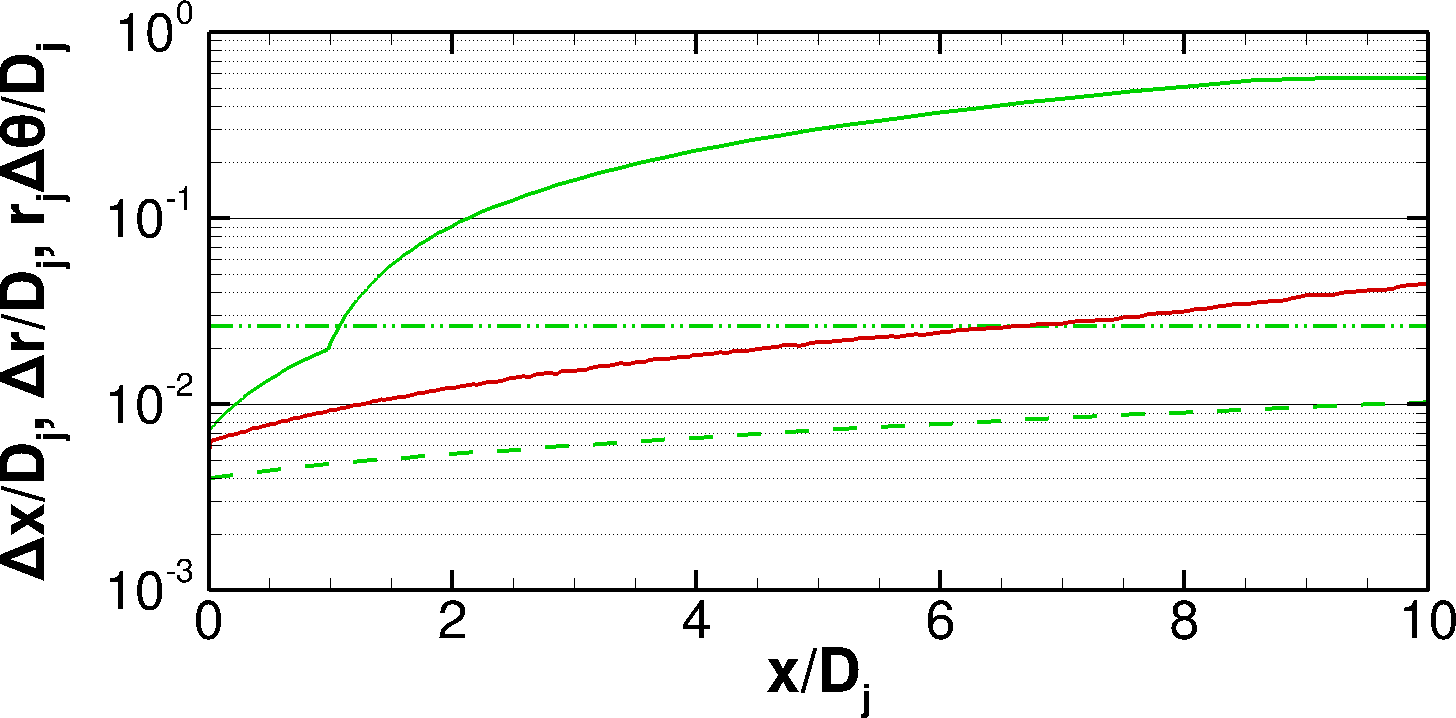}
		\end{minipage}
		\hfill
		\begin{minipage}[c]{\linewidth}
			\vspace{0.1cm}
	 		\caption[]{Comparison of the longitudinal evolution of the mesh size at $r/D_j = 0.5$ for the \emph{Aghora} simulations. \textcolor{red}{\trait}, tetrahedral mesh; \textcolor{green}{\trait}, hexahedral mesh; \trait, $\Delta x$; \tirets, $\Delta r$; \dmixte, $r\Delta\theta$}
 			\label{fig_taille_maille_hexa}
		\end{minipage}	
	\end{center}
\vspace{-0.8cm}
\end{figure}

Figure~\ref{fig_taille_maille_hexa} represents the longitudinal evolutions of the mesh size along a mesh line at $r/D_j = 0.5$ for the hexahedral mesh, as well as the tetrahedral mesh used for the simulation performed with \emph{Aghora}. At the nozzle lip, the axial mesh size is $\Delta x/D_j = 0.72\%$ and is then stretched with a rate of $1.3\%$ up to $x/D_j = 1$ and then at a rate of $7\%$. With these parameters the maximum axial mesh size is reached around $x/D_j \sim 9$, i.e.\ downstream of the estimated location of the potential core end. For $x/D_j \geq 9$, the axial mesh size is constant up to the downstream end of the refined mesh zone, i.e.\ $x/D_j = 35$. Finally a stretching of $20\%$ is applied up to the downstream boundary of the computational domain to avoid spurious reflections.  The 3D grid has been built by rotation of a 2D grid, except around the axis for which an "O-H" grid type has been used. 120 azimuthal planes have been used for this mesh, it allows for a moderate anisotropy near the nozzle lip, i.e.\ at the beginning of the shear layer. The mesh size is greater for the hexahedral mesh in this zone than the tetrahedral mesh. However we have seen (see section~\ref{subsubsec_TGV}) that the hexahedral mesh associated with a full tensored basis leads to lower levels of dissipation in the solution. All these characteristics are summarized in table~\ref{tab_mesh_charac_hexa}.
\\

\begin{table}[h]
 	\begin{center}
    	\begin{tabular}{l*{6}{c}}
  		\hline \hline
      Grid name				& $\Delta/D_j$ (\%)	& $n_\theta$	& $St_{cut-off} $	& \#cells ($\times 10^6$)	& \#DoFs ($\times 10^6$)	& $T.U_j/D_j$	\\ \hline
       TetraP3					& 0.8						& 390				& 1.5						& 3.9									& 78											& 180 				\\  \hline
      HexaP3\_VMS		& 0.72; 0.4; 2.6		& 120				& 1.0						& 3.2									& 208										& 45					\\  \hline
  		\end{tabular}
		\vspace{0.1cm}
    	\caption{Characteristics of the different grids, the mesh size $\Delta$ ($\Delta x$, $\Delta r$ and $r\Delta\theta$ for HexaP3\_VMS) and the number of cells in the azimuthal direction $n_\theta$ are given near $x/D_j = 0$ and $r/D_j = 0.5$. $T$ represents the simulated duration.}
     	\label{tab_mesh_charac_hexa}
 	\end{center}
	\vspace{-0.5cm}
\end{table}


\subsubsection{Shear layer}
\label{subsubsec_ShearLayer_Hexa}

%

\begin{figure}[h!]
	\begin{center}
		\begin{minipage}[c]{0.49\linewidth}
			\centering \includegraphics[height=3.5cm]{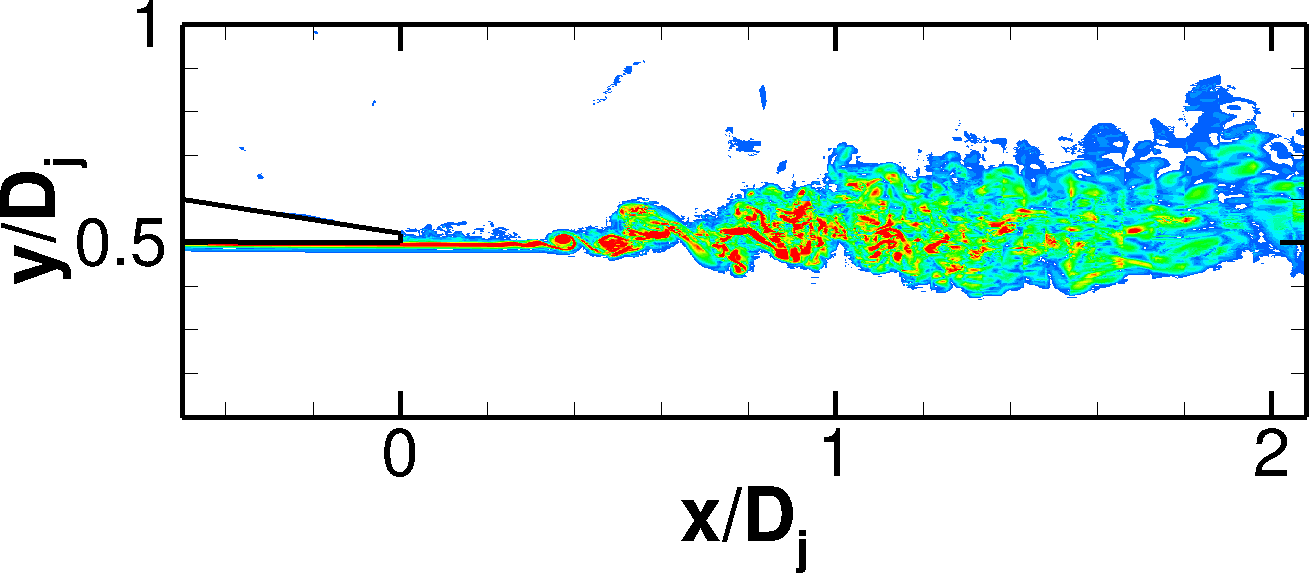}
		\end{minipage}
		\hfill
		\begin{minipage}[c]{0.49\linewidth}
			\centering \includegraphics[height=3.5cm]{./Vort_ProjSortieTuyere_aghora_TetraP3_r70.png}
		\end{minipage}
		\hfill
		\vspace{0.2cm}
		\begin{minipage}[c]{0.49\linewidth}
			\centering (b) HexaP3\_VMS
		\end{minipage}
		\hfill
		\vspace{0.2cm}
		\begin{minipage}[c]{0.49\linewidth}
			\centering (a) TetraP3
		\end{minipage}
		\hfill
		\begin{minipage}[c]{\linewidth}
			\vspace{0.1cm}
	 		\caption[]{Snapshots in the (x,r) plane of the vorticity norm in the boundary layer and the shear layer downstream of the lip line for $0 \leq  |\omega| \leq 60U_j/D_j$}
 			\label{fig_vort_sortie_tuyere_COMP_TetraP3_HexaP3VMS}
		\end{minipage}	
	\end{center}
\vspace{-0.8cm}
\end{figure}

\hspace{0.5cm} Figure~\ref{fig_vort_sortie_tuyere_COMP_TetraP3_HexaP3VMS} compares the instantaneous vorticity fields in the shear layer for the TetraP3 simulation (presented in section~\ref{sec_Valid}) and the HexaP3\_VMS simulation. Looking at these two vorticity fields, we can see that, similarly to the TetraP3 simulation, the HexaP3\_VMS simulation has an initially laminar shear layer with higher vorticity levels and that seems to transition to a turbulent state a bit further from the nozzle exit than the TetraP3 shear layer ($x/D_j \approx 0.5$ against $x/D_j \approx 0.4$). Moreover the HexaP3\_VMS simulation exhibits a thicker shear layer which seems to develop faster than for the TetraP3 simulation.
\\

%

\begin{figure}[h!]
	\begin{center}
		\begin{minipage}[c]{\linewidth}
			\centering \includegraphics[height=4cm]{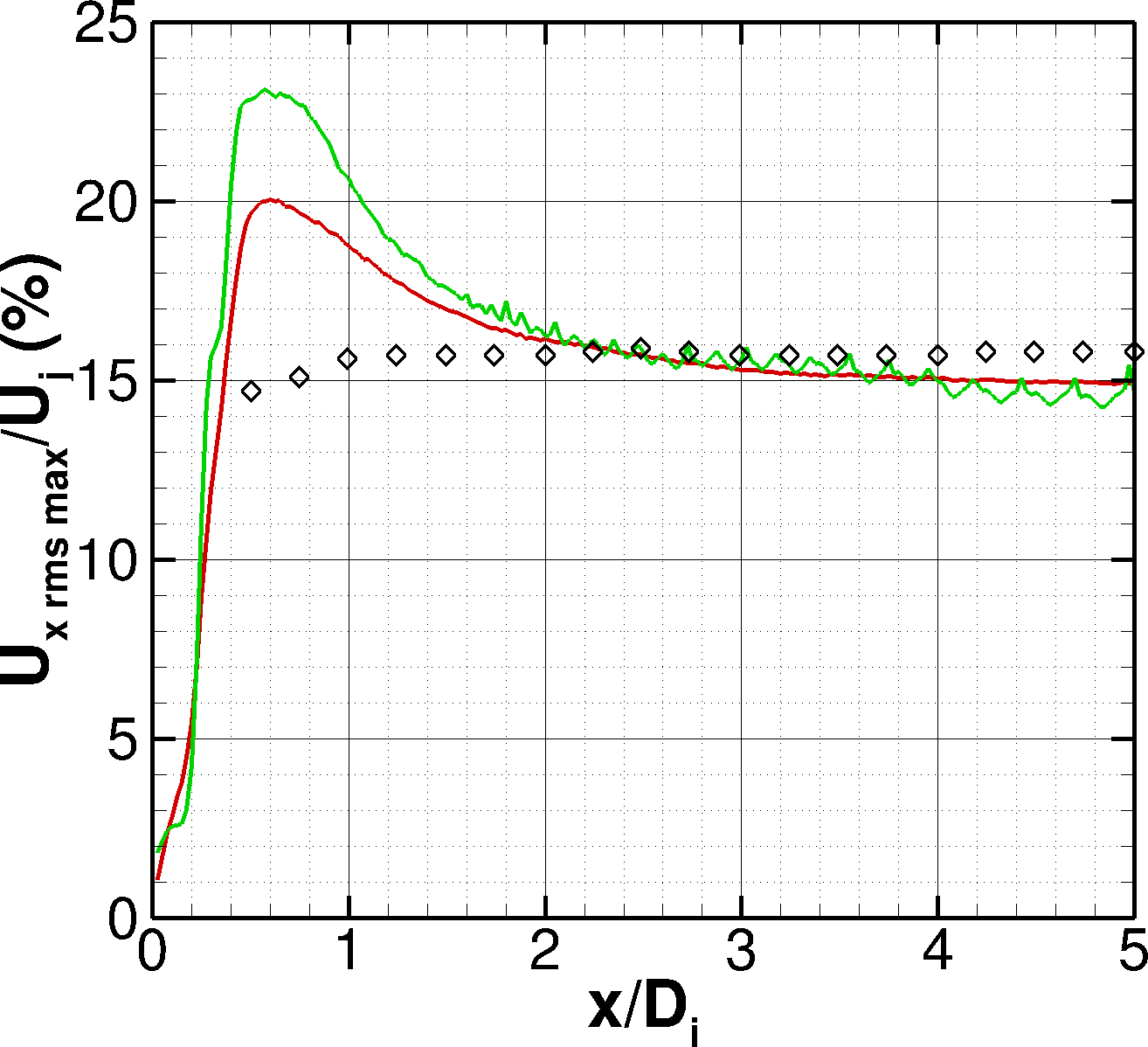}
		\end{minipage}
		\hfill
		\begin{minipage}[c]{\linewidth}
			\vspace{0.1cm}
	 		\caption[]{Comparison of the longitudinal evolution of the peak rms value of axial velocity. \textcolor{red}{\trait}, TetraP3; \textcolor{green}{\trait}, HexaP3\_VMS; $\diamond$, Fleury \emph{et al.} ($M=0.9$, $Re=7.7\times10^5$)}
 			\label{fig_Ux_rms_max_TetraP3_HexaFineP3VMS}
		\end{minipage}	
	\end{center}
\vspace{-0.8cm}
\end{figure}

To better quantify these differences, we can look at the longitudinal evolutions  of the rms of the axial velocity $U_{x rms}$ presented in figure~\ref{fig_Ux_rms_max_TetraP3_HexaFineP3VMS}. The HexaP3\_VMS simulation presents a similar growth of the $U_{x rms}$ peak value from the nozzle exit but reaches a higher value at around the same position ($23\%$ against $20\%$ at $x/D_j \approx 0.6$) with a sharper peak for the TetraP3. The $U_{x rms}$ levels then decrease to reach a value of around $15-16\%$ for $x/D_j \geq 2$. These higher turbulence levels may be related to the lower dissipation introduced by the VMS approach.
\\

\begin{figure}[h!]
	\begin{center}
		\begin{minipage}[c]{0.32\linewidth}
			\centering \includegraphics[height=4cm]{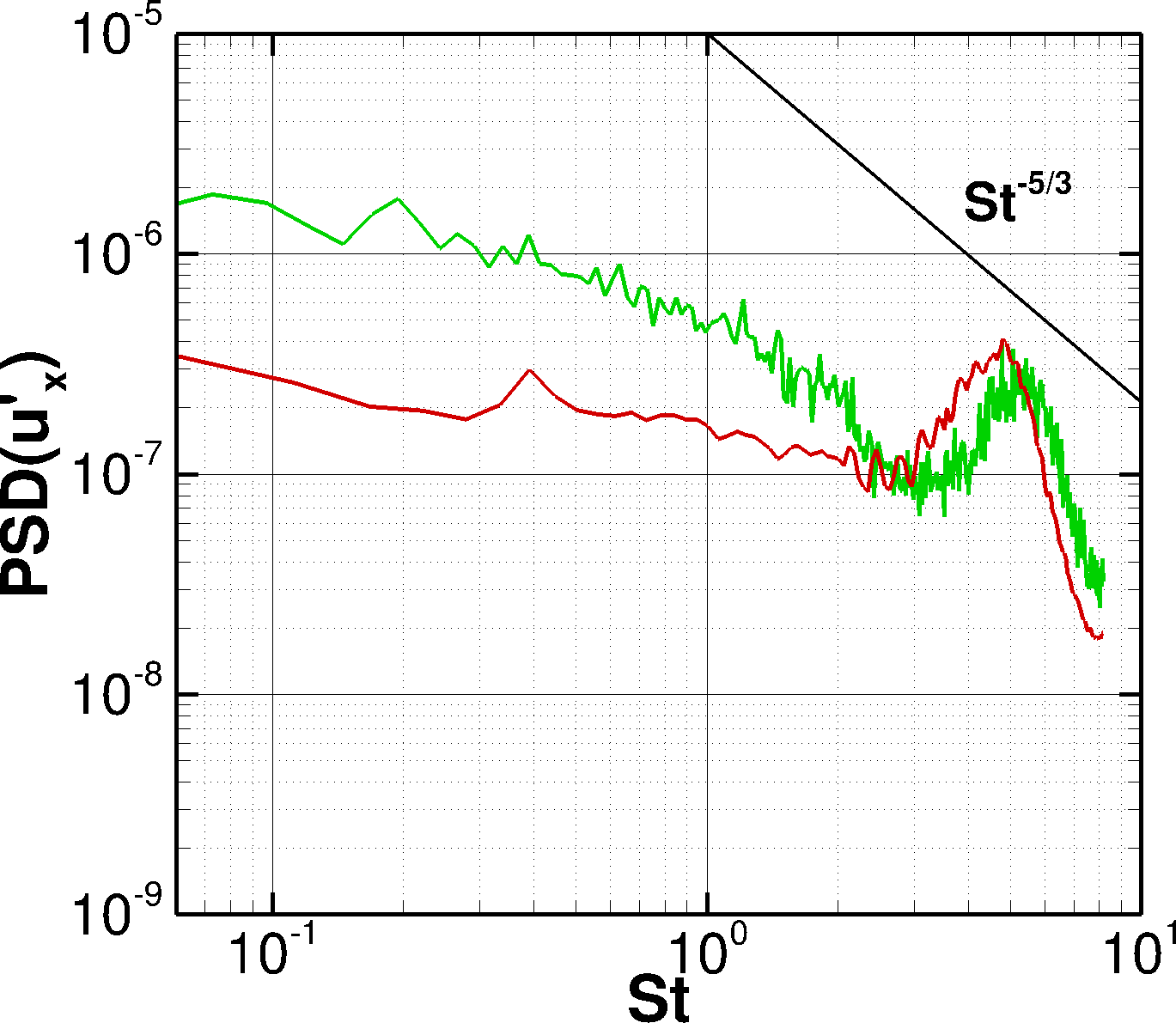}
		\end{minipage}
		\hfill
		\begin{minipage}[c]{0.32\linewidth}
			\centering \includegraphics[height=4cm]{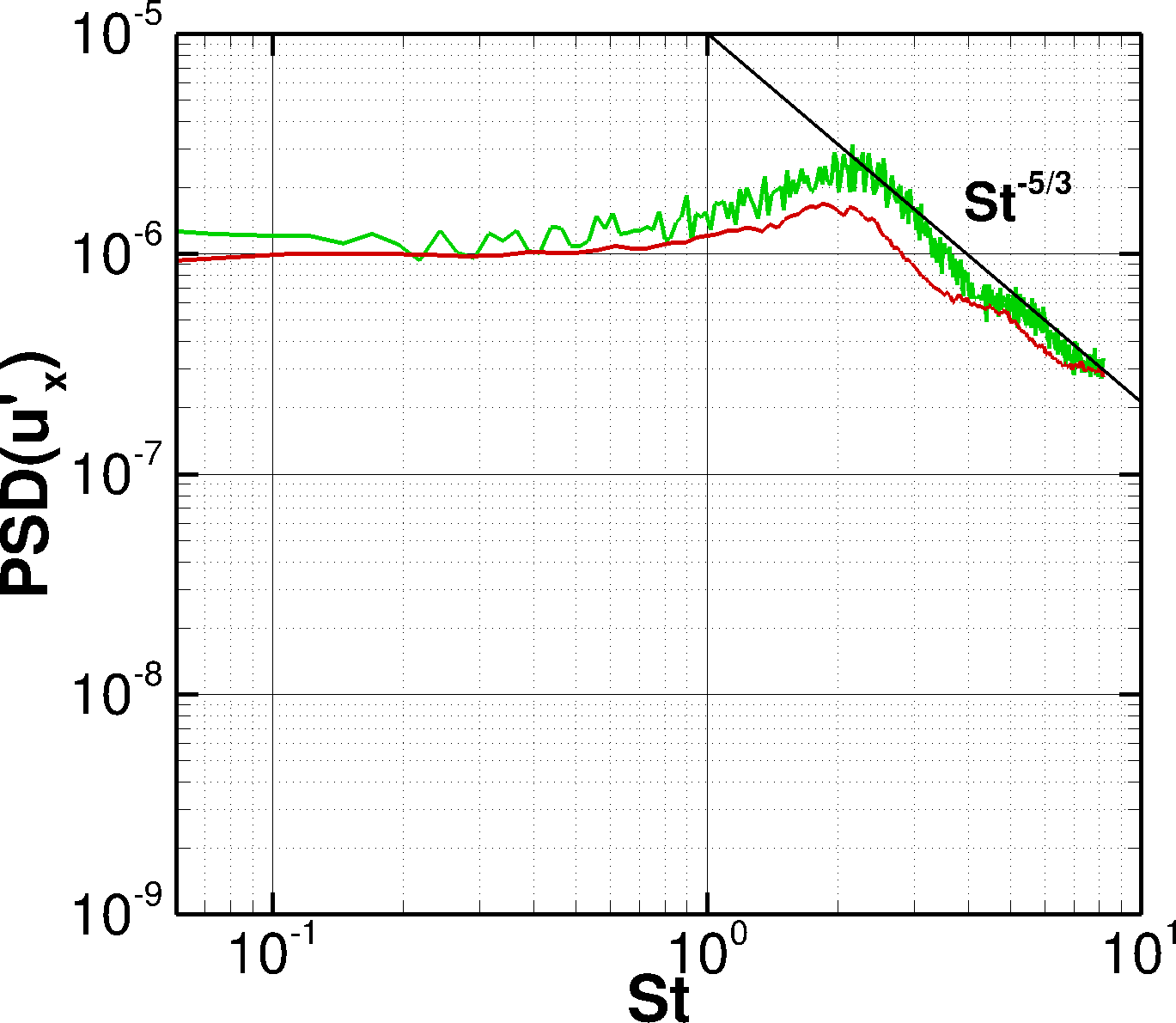}
		\end{minipage}
		\hfill
		\begin{minipage}[c]{0.32\linewidth}
			\centering \includegraphics[height=4cm]{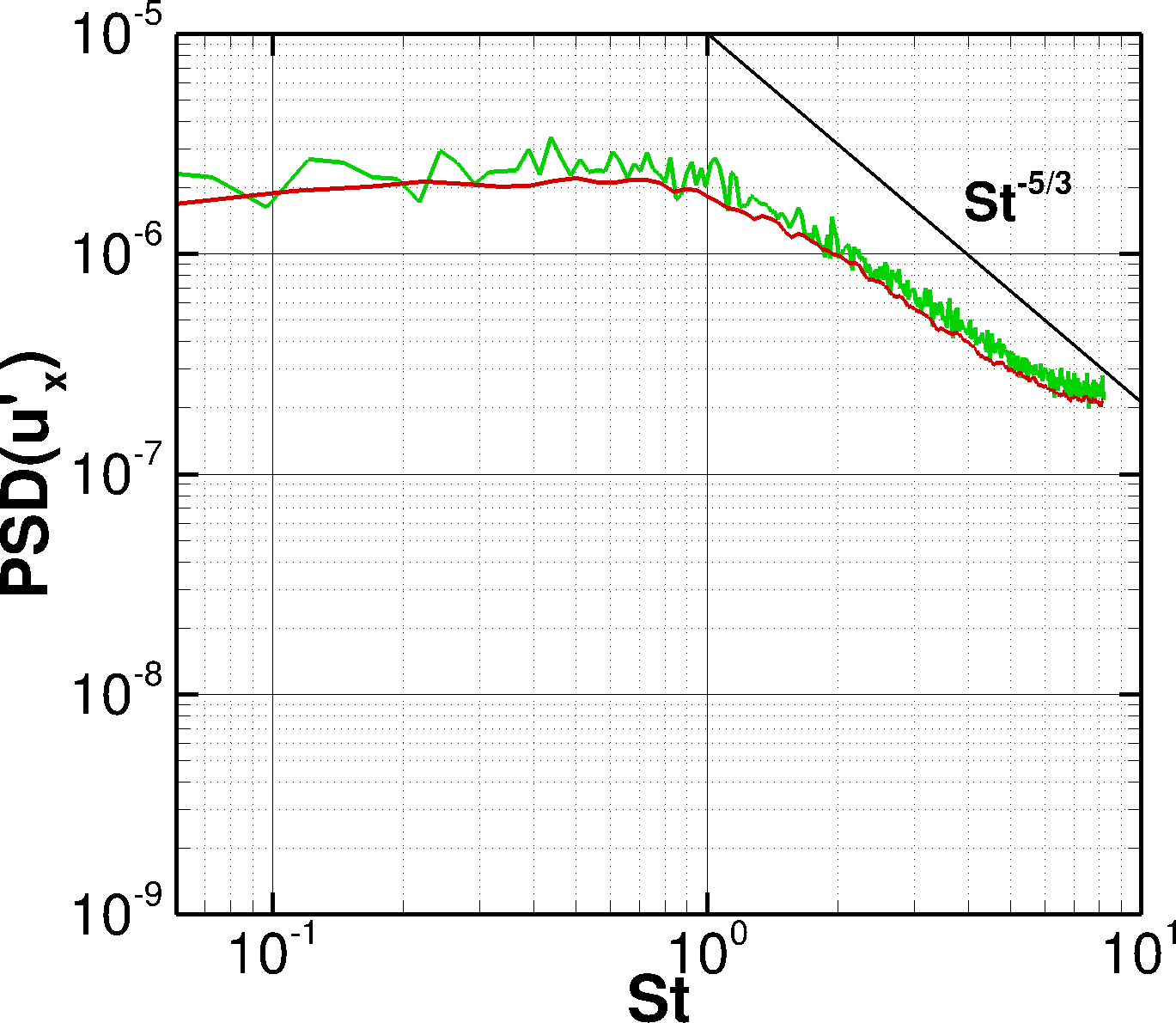}
		\end{minipage}
		\hfill
		\vspace{0.1cm}
		\begin{minipage}[c]{0.32\linewidth}
			\centering (a) $x/D_j = 0.25$
		\end{minipage}
		\hfill
		\vspace{0.1cm}
		\begin{minipage}[c]{0.32\linewidth}
			\centering (b) $x/D_j = 0.5$
		\end{minipage}
		\hfill
		\vspace{0.1cm}
		\begin{minipage}[c]{0.32\linewidth}
			\centering (c) $x/D_j = 1$
		\end{minipage}
		\hfill
		\begin{minipage}[c]{\linewidth}
			\vspace{0.1cm}
	 		\caption[]{Comparison of PSD of axial fluctuating velocity adimensioned by $U_j$ at $r/D_j = 0.5$ for $x/D_j \in \{0.25; 0.5; 1\}$. \textcolor{red}{\trait}, TetraP3; \textcolor{green}{\trait}, HexaP3\_VMS}
 			\label{fig_DSP_Ux_cisail_COMP_TetraP3_HexaFineP3VMS}
		\end{minipage}	
	\end{center}
\vspace{-0.8cm}
\end{figure}

In figure~\ref{fig_DSP_Ux_cisail_COMP_TetraP3_HexaFineP3VMS}, depicting the DSP of the fluctuating axial velocity in the shear layer, we can see that the stronger growth of the $U_{x rms}$ levels of the HexaP3\_VMS are related to the low frequency part of the spectra. These higher low frequencies levels might be due to the VMS approach. Indeed, with the VMS approach, the dissipation is concentrated on the smaller scales contrary to the LES approach for which all scales are affected. However downstream of the peak value, i.e.\ $x/D_j \geq 0.5$,  the spectra are similar. As discussed in section~\ref{subsec_ShearLayer}, we can also see on the PSDs several high frequency peaks at $St \approx 4$ for $x/D_j = 0.25$ and $St \approx 2$ for $x/D_j = 0.5$, which are related to pairings occurring in the shear layer. And sufficiently downstream of the nozzle exit, e.g.\ $x/D_j \geq 1$, these peaks are not present anymore.
\\

\begin{figure}[h!]
	\begin{center}
		\begin{minipage}[c]{0.32\linewidth}
			\centering \includegraphics[height=4cm]{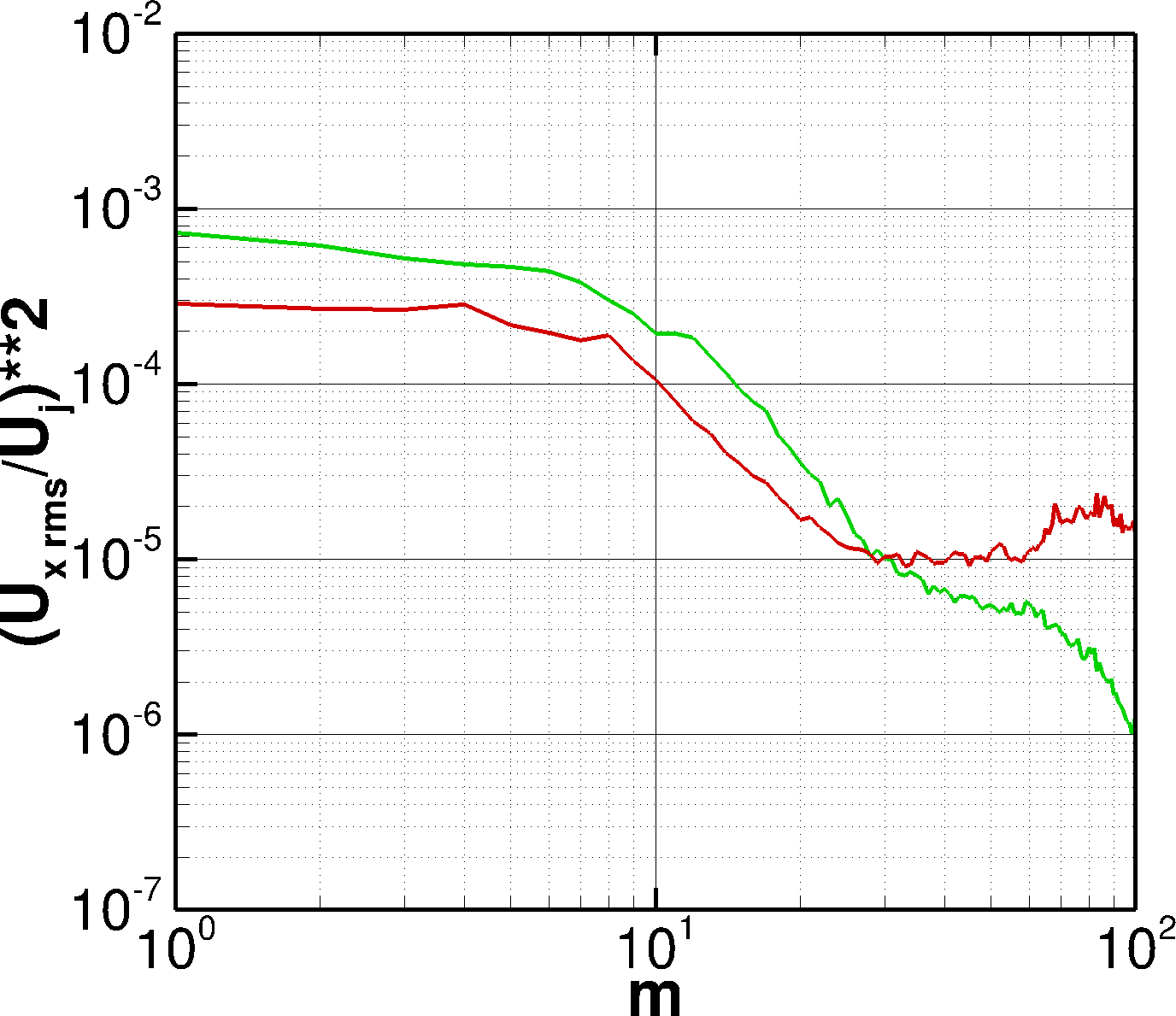}
		\end{minipage}
		\hfill
		\begin{minipage}[c]{0.32\linewidth}
			\centering \includegraphics[height=4cm]{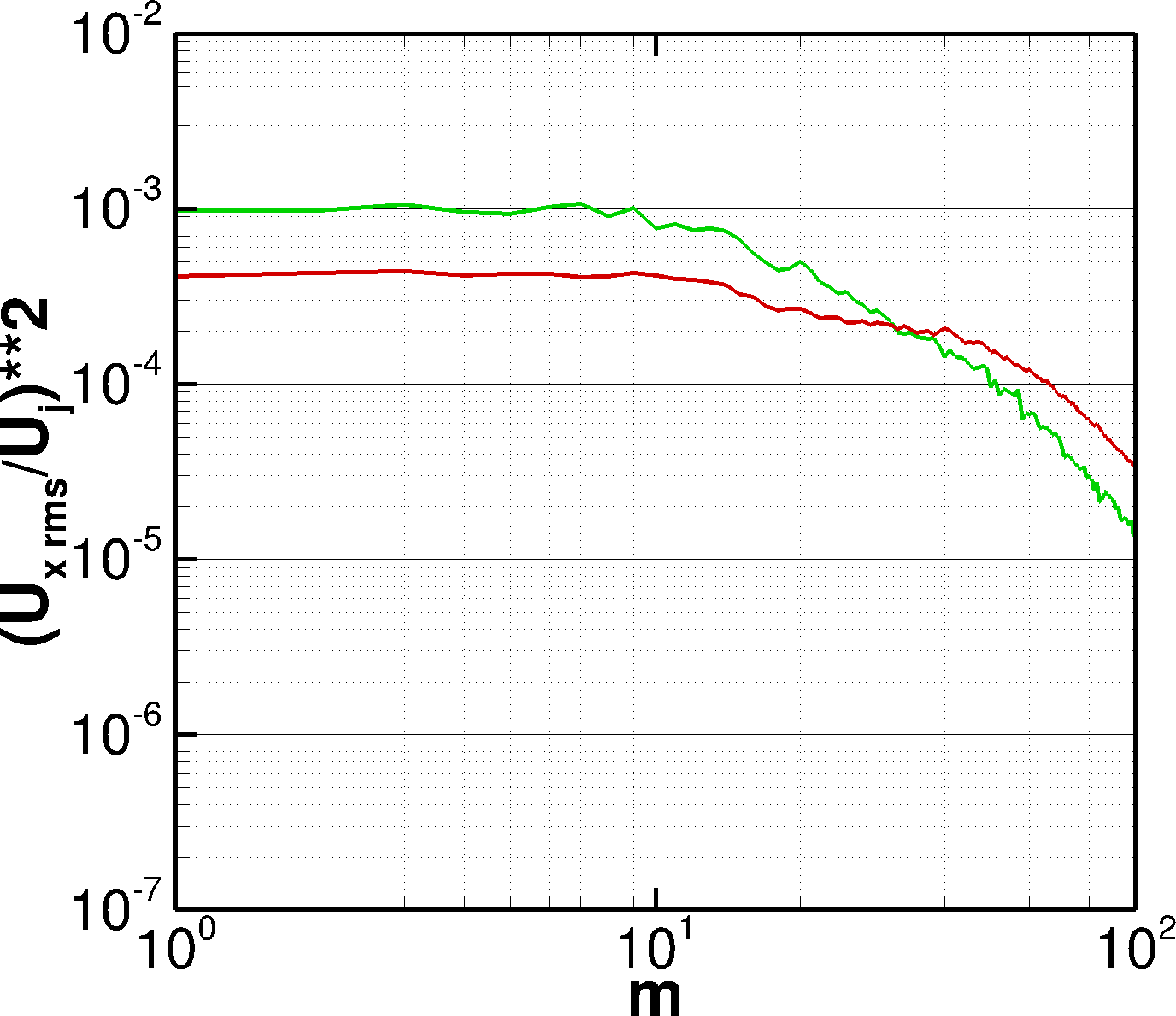}
		\end{minipage}
		\hfill
		\begin{minipage}[c]{0.32\linewidth}
			\centering \includegraphics[height=4cm]{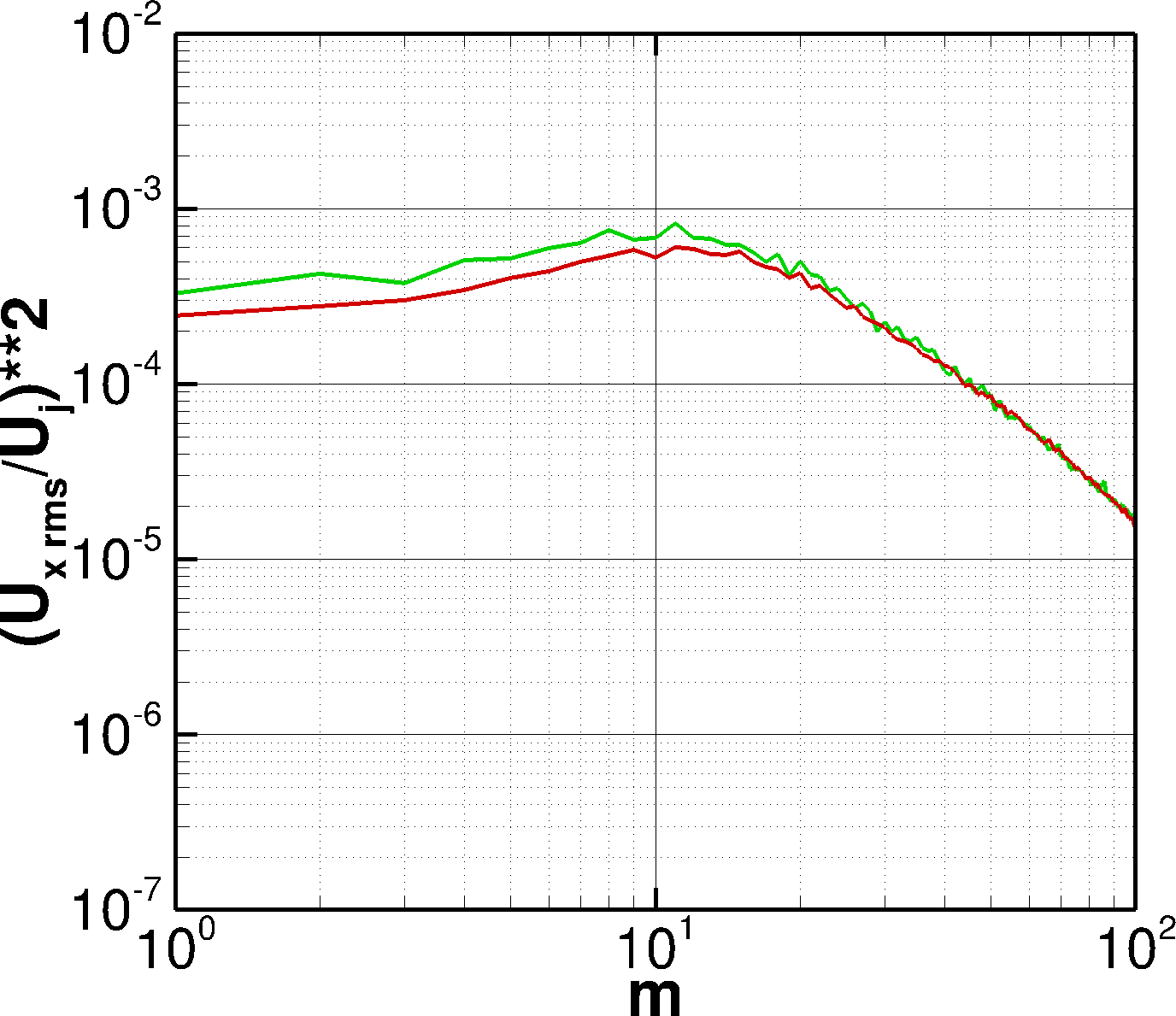}
		\end{minipage}
		\hfill
		\vspace{0.1cm}
		\begin{minipage}[c]{0.32\linewidth}
			\centering (a) $x/D_j = 0.25$
		\end{minipage}
		\hfill
		\vspace{0.1cm}
		\begin{minipage}[c]{0.32\linewidth}
			\centering (b) $x/D_j = 0.5$
		\end{minipage}
		\hfill
		\vspace{0.1cm}
		\begin{minipage}[c]{0.32\linewidth}
			\centering (c) $x/D_j = 1$
		\end{minipage}
		\hfill
		\begin{minipage}[c]{\linewidth}
			\vspace{0.1cm}
	 		\caption[]{Comparison of rms axial fluctuating velocity adimensioned by $U_j$ at $r/D_j = 0.5$ for $x/D_j \in \{0.25; 0.5; 1\}$ as function of the azimuthal mode order $m$. \textcolor{red}{\trait}, TetraP3; \textcolor{green}{\trait}, HexaP3\_VMS}
 			\label{fig_ModesAzim_Ux_cisail_COMP_TetraP3_HexaFineP3VMS}
		\end{minipage}	
	\end{center}
\vspace{-0.8cm}
\end{figure}

Similarly to the DSP shown in figure~\ref{fig_DSP_Ux_cisail_COMP_TetraP3_HexaFineP3VMS}, the azimuthal composition of the fluctuating axial velocity presented in figure~\ref{fig_ModesAzim_Ux_cisail_COMP_TetraP3_HexaFineP3VMS} presents higher levels for the lower order modes for the HexaP3\_VMS simulation compared to the TetraP3 one, at position close to the nozzle exit $x/D_j \leq 0.5$. At downstream positions, the azimuthal compositions are similar, with low order azimuthal $m \leq 10$ more energetic for the HexaP3\_VMS simulation. It might be related to the faster development of the jet flow mentioned previously, large scales and thus low order azimuthal modes grow faster for the HexaP3\_VMS simulation.
\\

We have seen in this section that, similarly to the TetraP3 simulation, the HexaP3\_VMS simulation exhibits a shear layer initially laminar which transitions downstream of the nozzle exit but which is more energetic especially for the low frequencies and the low-order azimuthal modes. This might be related to the VMS approach through which the dissipation associated with the Smagorinsky subgrid-scale model is concentrated at the small-scales level and thus the large scales are not directly affected. We will now study how these different behaviours affect the jet development.

\subsubsection{Jet development}
\label{subsubsec_JetDev_Hexa}

\begin{figure}[h!]
	\begin{center}
		\begin{minipage}[c]{0.49\linewidth}
			\centering \includegraphics[height=4cm]{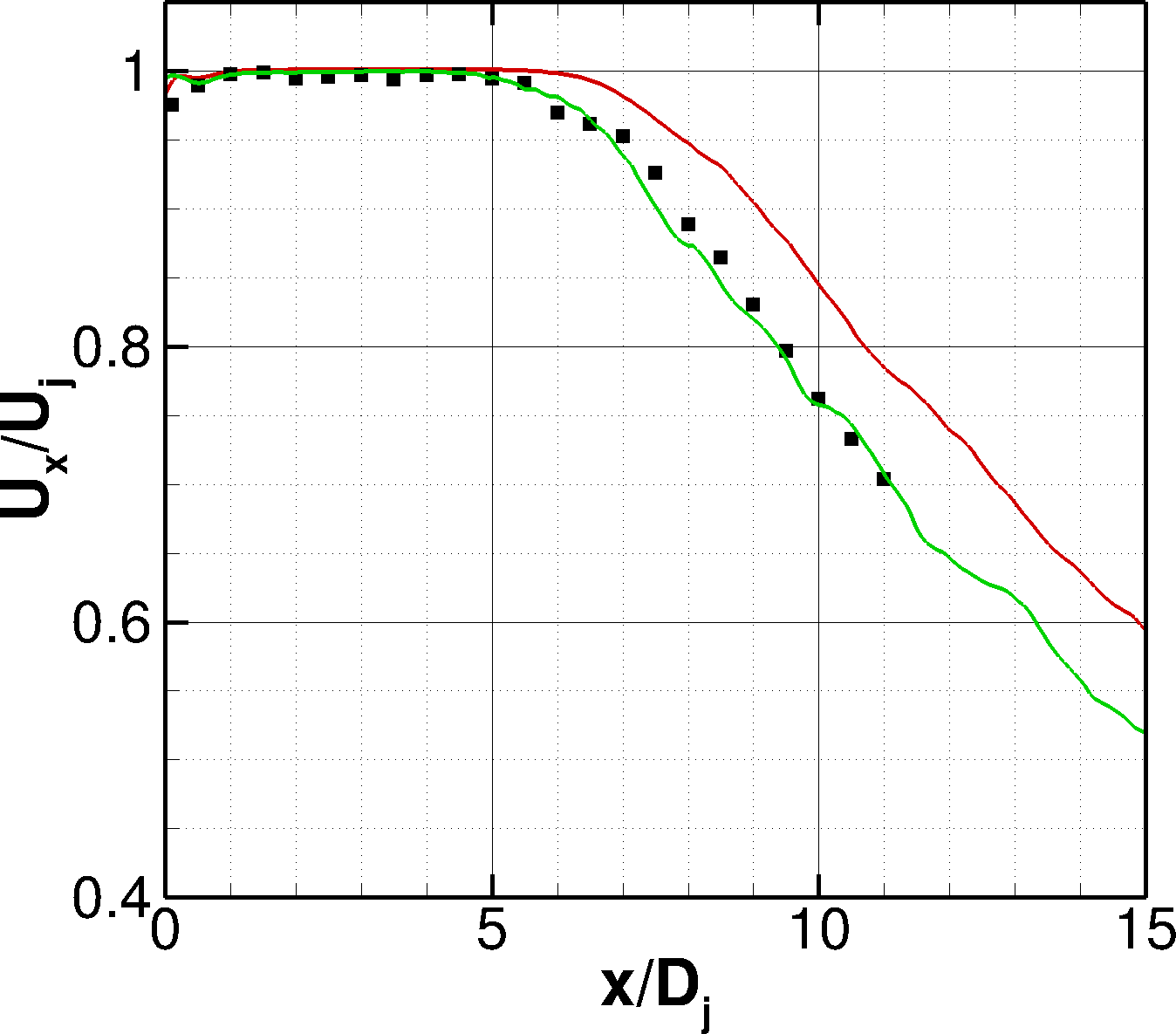}
		\end{minipage}
		\hfill
		\begin{minipage}[c]{0.49\linewidth}
			\centering \includegraphics[height=4cm]{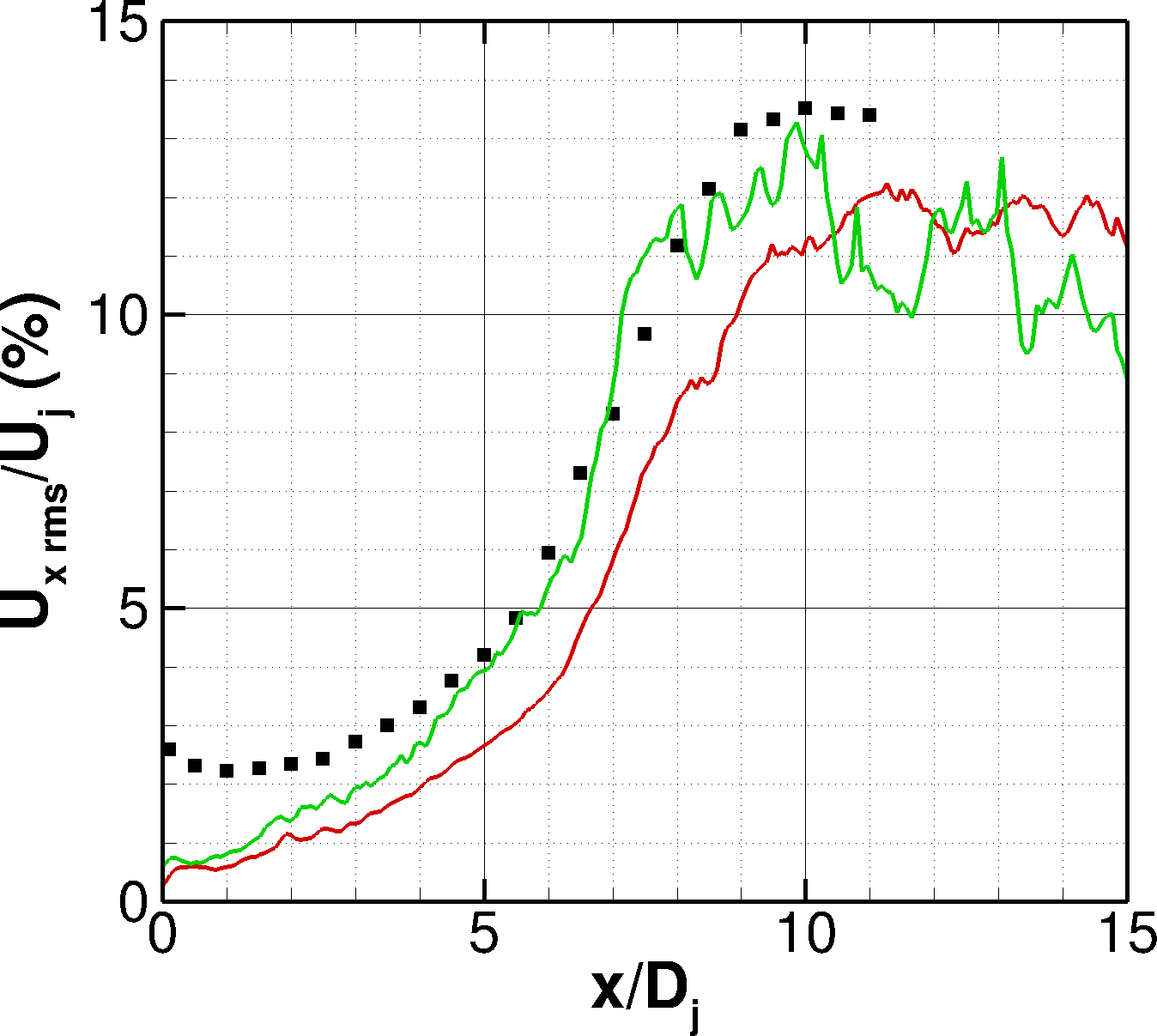}
		\end{minipage}
		\hfill
		\vspace{0.1cm}
		\begin{minipage}[c]{0.49\linewidth}
			\centering (a)
		\end{minipage}
		\hfill
		\vspace{0.1cm}
		\begin{minipage}[c]{0.49\linewidth}
			\centering (b)
		\end{minipage}
		\hfill
		\begin{minipage}[c]{\linewidth}
			\vspace{0.1cm}
	 		\caption[]{Comparison of the axial evolution on the jet axis of the (a) mean and (b) rms axial velocity. \textcolor{red}{\trait}, TetraP3; \textcolor{green}{\trait}, HexaP3\_VMS;  $\blacksquare$, measurements}
 			\label{fig_Ux_moy_rms_axe_COMP_TetraP3_HexaFineP3VMS}
		\end{minipage}	
	\end{center}
\vspace{-0.8cm}
\end{figure}

\hspace{0.5cm} Figure~\ref{fig_Ux_moy_rms_axe_COMP_TetraP3_HexaFineP3VMS} presents the longitudinal evolutions of the mean and rms axial velocity on the jet axis for the two DG simulations performed. The differences highlighted through the analysis of the shear layer state close to the nozzle exit (i.e.\ $x/D_j \leq 1$) are also visible on the jet flow development. Indeed, the three simulations have different potential core lengths. The HexaP3\_VMS simulation gives $L_c/D_j \approx 6.83$. As a reminder, the experimental data give $L_c/D_j = 7$ and the TetraP3 simulation $L_c/D_j \approx 7.9$. The shorter potential core obtained with the HexaP3\_VMS simulation is a confirmation of the faster development of the shear layer observed in the previous section. 

When looking at the axial velocity rms value along the jet axis presented in figure~\ref{fig_Ux_moy_rms_axe_COMP_TetraP3_HexaFineP3VMS}(b), we can see that the levels are not sufficiently converged for the hexahedral simulations as the simulated duration is rather short compared to the TetraP3 simulation (see table \ref{tab_mesh_charac_hexa}). However some tendencies can be observed. Related to the higher spreading rate and higher peak rms velocity of its shear layer compared to the TetraP3 one, the HexaP3\_VMS simulation presents rms levels on the jet axis growing faster and reaching higher values.
\\

We have seen that the HexaP3\_VMS simulation exhibits some differences in the shear layer state close to the nozzle exit (thicker shear layer, higher peak value of the $U_{x rms}$, more energetic low order modes) when compared to the TetraP3 simulation, despite both simulations presenting an initially laminar shear layer. This results in a faster development of the shear layer and thus a shorter potential. As mentioned in section~\ref{subsec_JetDev}, Bogey \& Bailly~\cite{Bogey2010} have shown that the differences in behaviour can be linked to the different initial shear layer thickness of the simulations. Indeed, they have shown that for an initially laminar jet, the thinner the shear layer is, the sooner the shear layer develops but at a slower pace. Another effect at play might come from the dissipation introduced by the Smagorinsky subgrid-scale model used for the present simulations. It might affect the boundary layer which might affect in turn the initial development of the shear layer which plays an important role in the jet development and its radiated pressure field.
Nonetheless, the two simulations performed with the DG method do not reproduce well enough the experimental data. Some form of turbulence triggering might be needed in order to correctly compute the jet flow as shown by Bogey \emph{et al.}~\cite{Bogey2012}.

\section{Conclusions}



\hspace{0.5cm} The present work deals with the numerical simulation of the aerodynamic and acoustic fields of an isothermal subsonic turbulent jet using LES in combination with a high-order DG method. 

Firstly, a simulation has been performed using a DG-P3 approximation ($4^{th}$-order accuracy in space) on a fully unstructured tetrahedral mesh and compared to a reference simulation carried out using a finite volume method on a similar (finer) mesh. The results have been compared with experimental data for the flow and acoustic fields.

The DG simulation presents a laminar shear layer at the nozzle exit which transitions to turbulence at $x/D_j \approx 0.4$ instead of being fully turbulent right after the nozzle exit, as is usually the case in simulated jets. This suggests that some form of turbulence triggering might be necessary. This transition probably associated with other effects, such as a too thin boundary layer or Reynolds number effects due to the subgrid-scale model, leads to an overestimation of the potential core length and an underestimated rms level of the axial velocity on the jet axis. This in turn leads to an underestimation of the low frequency content of the downstream acoustic spectra in the far field. 

%

In a second stage, we have performed another simulation using a hexahedral mesh associated with a full tensored set of basis functions as opposed to a Dubiner set of basis functions as is the case for a tetrahedron, with the same spatial and temporal discretizations. The VMS approach has also been used in order to limit the dissipation introduced by the Smagorinsky subgrid-scale model. The two DG simulations results have been compared for the shear layer and flow field development. Similarly to the simulation performed on the tetrahedral mesh, the simulation performed on the hexahedral mesh presents an initially laminar shear layer which transitions to a turbulent state around the same position but with higher energy levels at low frequencies and low azimuthal mode orders. This might be due to the smaller amount of dissipation present in the VMS approach. This more energetic transition might cause a shortening of the potential core of the hexahedral simulation compared to the tetrahedral simulation. 

To improve the results and reduce the computational cost of the simulation (compared to a $5^{th}$-order simulation), local "p" adaptation will also be used (in particular by using a higher p value close to the nozzle exit).

\section*{Acknowledgments}

\hspace{0.5cm} This work was performed using HPC resources from GENCI (Grant 2016-c2016067561). The authors would like to thank Dr. Emeric Martin and Dr. F. Renac for their valuable help with the parallel computing issues and on the numerical methods, M.-C. Le Pape and F. Vuillot for their help in high-order mesh generation process and N. Lupoglazoff for providing the finite volume data. This work has been supported by the PIA ELCI project of the French FSN.


\bibliographystyle{unsrt}
\bibliography{biblio.bib}

\begin{thebibliography}{10}

\bibitem{Husain1979}
Z.~D. Husain and A.~K. M.~F. Hussain.
\newblock {Axisymmetric mixing layer: influence of the initial and boundary
  conditions}.
\newblock {\em AIAA Journal}, 17(1), 1979.

\bibitem{Zaman1985_1}
K.~B. M.~Q. Zaman.
\newblock {Effect of initial condition on subsonic jet noise}.
\newblock {\em AIAA Journal}, 23(9), 1985.

\bibitem{Huet2013}
M.~Huet.
\newblock {Influence of boundary layers resolution on heated, subsonic, high
  Reynolds number jet flow and noise}.
\newblock In {\em {AIAA paper}}, {(2141)}, 2013.

\bibitem{Bogey2011}
C.~Bogey, O.~Marsden, and C.~Bailly.
\newblock {Large-eddy simulation of the flow and acoustics fields of a Reynolds
  number $10^5$ subsonic jet with tripped exit boundary layers}.
\newblock {\em Physics of Fluids}, 23, 2011.

\bibitem{Bogey2012}
C.~Bogey, O.~Marsden, and C.~Bailly.
\newblock {Influence of initial turbulence level on the flow and sound fields
  of a subsonic jet at a diameter-based Reynolds number of $10^5$}.
\newblock {\em Journal of Fluid Mechanics}, 701, 2012.

\bibitem{Zaman1985_2}
K.~B. M.~Q. Zaman.
\newblock {Far-field noise of a subsonic jet under controlled excitation}.
\newblock {\em Journal of Fluid Mechanics}, 152, 1985.

\bibitem{Bogey2003}
C.~Bogey, C.~Bailly, and D.~Juv{\'e}.
\newblock {Noise investigation of a high subsonic, moderate Reynolds number jet
  using a compressible large eddy simulaiton}.
\newblock {\em Theoretical and Computational Fluid Dynamics}, 16(4), 2003.

\bibitem{Bogey2005}
C.~Bogey and C.~Bailly.
\newblock {Effect of inflow conditions and forcing on subsonic jet flows and
  noise}.
\newblock {\em AIAA Journal}, 43(5), 2005.

\bibitem{Lew2005}
P.~Lew, G.~A. Blaisdell, and A.~S. Lyrintzis.
\newblock {Recent progress of hot jet aeroacoustics using 3-D Large-Eddy
  Simulation}.
\newblock In {\em {AIAA paper}}, No. 2005-3084, 2005.

\bibitem{Uzun2005}
A.~Uzun, A.~S. Lyrintzis, and G.~A. Blaisdell.
\newblock {Coupling of integral acoustics methods with LES for jet noise
  prediction}.
\newblock {\em International Journal of Aeroacoustics}, 3(4), 2005.

\bibitem{Zhao2001}
W.~Zhao, S.~H. Frankel, and L.~Mongeau.
\newblock {Large Eddy Simulation of sound radiation from subsonic turbulent
  jets}.
\newblock {\em AIAA Journal}, 39(8):1469--1477, 2001.

\bibitem{Bodony2005}
D.~J. Bodony and S.~K. Lele.
\newblock {On using large-eddy simulation for the prediction of noise from cold
  and heated turbulent jets}.
\newblock {\em Physics of Fluids}, 17, 2005.

\bibitem{Fosso2012}
A.~Fosso-Pouangu{\'e}, M.~Sanjos{\'e}, and S.~Moreau.
\newblock {Jet noise simulation with realistic nozzle geometries using fully
  unstructred LES solver}.
\newblock In {\em {AIAA paper}}, {(2190)}, 2012.

\bibitem{Sanjose2014}
M.~Sanjos{\'e}, A.~Fosso-Pouangu{\'e}, S.~Moreau, G.~Wang, and T.~Padois.
\newblock {Unstructures LES of baseline EXEJET dual-stream jet}.
\newblock In {\em {AIAA paper}}, {(3037)}, 2014.

\bibitem{Lorteau2015}
M.~Lorteau, F.~Cl{\'e}ro, and F.~Vuillot.
\newblock {Analysis of noise radiation mechanisms in hot subsonic jet from a
  validated LES solution}.
\newblock {\em Physics of Fluids}, 27, 2015.

\bibitem{Chapelier2014}
M.~Chapelier, J.-B.and de la Llave~Plata, F.~Renac, and E.~Lamballais.
\newblock {Evaluation of a high-order DG method for the DNS of turbulent
  flows}.
\newblock {\em Computers \& Fluids}, 95, 2014.

\bibitem{Renac2015}
F.~Renac, M.~de~la Llave~Plata, E.~Martin, J.-B. Chapelier, and C.~Couaillier.
\newblock {Aghora: A High-Order DG Solver for Turbulent Flow Simulations,
  IDIHOM: Industrialisation of High-Order Methods - A Top Down Approach}.
\newblock {\em Notes on Numerical Fluid Mechanics and Multidisciplinary
  Design}, 128, 2015.

\bibitem{Houston2005}
Paul Houston and Endre S{\"u}li.
\newblock A note on the design of hp-adaptive finite element methods for
  elliptic partial differential equations.
\newblock {\em Computer Methods in Applied Mechanics and Engineering},
  194(2):229--243, 2005.

\bibitem{Dolejvsi2013}
V{\'\i}t Dolej{\v{s}}{\'\i}.
\newblock {hp-DGFEM for nonlinear convection-diffusion problems}.
\newblock {\em Mathematics and Computers in Simulation}, 87:87--118, 2013.

\bibitem{Kuru2016}
G{\"o}kt{\"u}rk Kuru, Marta de~la Llave~Plata, Vincent Couaillier, R{\'e}mi
  Abgrall, and Fr{\'e}d{\'e}ric Coquel.
\newblock {An adaptive variational multiscale discontinuous Galerkin method for
  large eddy simulation}.
\newblock In {\em {AIAA paper}}, {(0584)}, 2016.

\bibitem{Ham2009}
Frank~E. Ham, Arjun Sharma, M.~Shoeybi, Sanjiva Lele, Parviz Moin, and
  E.~van~der Weide.
\newblock {Noise Prediction from Cold High-Speed Turbulent Jets Using
  Large-Eddy Simulation}.
\newblock In {\em {AIAA paper}}, {(0009)}, 2009.

\bibitem{Marek2015}
M.~Marek, A.~Tyliszczak, and A.~Boguslawski.
\newblock {Large eddy simulation of incompressible free round jet with
  discontinuous Galerkin method}.
\newblock {\em International Journal for Numerical Methods in Fluids}, 79,
  2015.

\bibitem{Jordan2002_1}
P.~Jordan, Y.~Gervais, J.-C. Valière, and H.~Foulon.
\newblock {Final results from single point measurements, Project delivrable
  D3.4, JEAN-EU 5th Framework Program, G4RD-CT2000-0313}.
\newblock Tech.rep., Laboratoire d'Etudes Aérodynamiques, 2002.

\bibitem{Jordan2002_2}
P.~Jordan, Y.~Gervais, J.-C. Valière, and H.~Foulon.
\newblock {Results from acoutic field measurements, Project delivrable D3.4,
  JEAN-EU 5th Framework Program, G4RD-CT2000-0313}.
\newblock Tech.rep., Laboratoire d'Etudes Aérodynamiques, 2002.

\bibitem{Lupoglazoff2015}
N.~Lupoglazoff and F.~Vuillot.
\newblock {Recent progress in numerical simulations for jet noise computation
  using LES on fully unstructured meshes}.
\newblock In {\em {AIAA paper}}, {(2369)}, 2015.

\bibitem{Vuillot2016}
F.~Vuillot, N.~Lupoglazoff, M.~Lorteau, and F.~Cl{\'e}ro.
\newblock {Large-Eddy Simulation of jet noise from unstructered grids with
  turbulent nozzle boundary layer}.
\newblock In {\em {AIAA paper}}, {(3046)}, 2016.

\bibitem{Remacle2003}
J.F. Remacle, J.E. Flaherty, and M.S. Shephard.
\newblock {An Adaptive Discontinuous Galerkin Technique with an Orthogonal
  Basis Applied to Compressible Flow Problems}.
\newblock {\em SIAM Review}, 45, 2003.

\bibitem{Bassi2012}
F.~Bassi, L.~Botti, A.~Colombo, D.A. Di~Pietro, and Tesini P.
\newblock {On the flexibility of agglomeration based physical space DG
  discretizations}.
\newblock {\em Journal of Computational Physics}, 231, 2012.

\bibitem{Gottlieb2001}
S.~Gottlieb, C.W. Shu, and E.~Tadmor.
\newblock {Strong stability-preserving high-order time discretization methods}.
\newblock {\em SIAM Review}, 43, 2001.

\bibitem{Arnold2001}
D.N. Arnold, F.~Brezzi, B.~Cockburn, and L.~D. Marini.
\newblock {Unified Analysis of Discontinuous Galerkin Methods for Elliptic
  Problems}.
\newblock {\em SIAM J. Numer. Anal.}, 39, 2001.

\bibitem{Fleury2008}
V.~Fleury, C.~Bailly, E.~Jondeau, M.~Michard, and D.~Juv{\'e}.
\newblock {Space-time correlations in two subsonic jets using dual particle
  image velocimetry measurements}.
\newblock {\em AIAA Journal}, 46(10), 2008.

\bibitem{Refloch2011}
A.~Refloch, B.~Courbet, A.~Murrone, P.~Villedieu, C.~Laurent, P.~Gilbank,
  J.~Troyes, L.~Tess\'e, G.~Chaineray, J.B. Dargaud, E.~Qu\'emerais, and
  Vuillot F.
\newblock {CEDRE software}.
\newblock {\em Aerospace Lab}, 2, 2011.

\bibitem{FWH1969}
J.~E. Ffowcs~Williams and D~.L. Hawkings.
\newblock {Sound generation by turbulence and surfaces in arbitrary motion}.
\newblock {\em Philosophical Transactions of the Royal Society of London A},
  264(1151), 1969.

\bibitem{Rahier2004}
G.~Rahier, J.~Prieur, F.~Vuillot, N.~Lupoglazoff, and A.~Biancherin.
\newblock {Investigation of integral surface formulations for acoustic
  post-processing of unsteady aerodynamic jet simulations}.
\newblock {\em Aerospace Science and Technology}, 8, 2004.

\bibitem{Bogey2010}
C.~Bogey and C.~Bailly.
\newblock {Influence of nozzle-exit boundary-layer conditions on the flow and
  acoustics fields of initially laminar jets}.
\newblock {\em Journal of Fluid Mechanics}, 663, 2010.

\bibitem{Tam2008}
C.~K.~W. Tam, K.~Viswanathan, K.~K. Ahuja, and J.~Panda.
\newblock {The sources of jet noise: experimental evidence}.
\newblock {\em Journal of Fluid Mechanics}, 615, 2008.

\bibitem{Taylor1937}
G.~E. Taylor and A.~E. Green.
\newblock {Mechanism of the production of small eddies from large ones}.
\newblock {\em Proceedings of the Royal Society of London. Series A.
  Mathematical and Physical Sciences}, 158(895), 1937.

\bibitem{Couaillier2005}
Vincent Couaillier.
\newblock {Effective Multidimensional Non Reflective Boundary Condition for CFD
  Calculations Applied to Turboengine Aeroacoustics Prediction}.
\newblock In {\em {ISABE paper}}, {(1185)}, 2005.

\end{thebibliography}

\end{document}